\documentclass[aps,twocolumn,nofootinbib]{revtex4}
\usepackage{graphicx}
\usepackage{epsfig}
\usepackage{amsmath}
\usepackage{amsfonts}

\begin{document}
\title{Physics Based on Physical Monism}
\author{Kim, Seong Dong}
\email[]{kimseongdong@paran.com} 
\date{\today} 
\begin{abstract}
Based on a physical monism, which holds that the matter and space
are classified by not a difference of their kind but a difference of
magnitude of their density, I derive the most fundamental equation
of motion, which is capable of providing a deeper physical
understanding than the known physics. For example, this equation
answers to the substantive reason of movement, and Newton's second
law, which has been regarded as the definition of force, is derived
in a substantive level from this equation. Further, the relativistic
energy-mass formula is generalized to include the potential energy
term, and the Lorentz force and Maxwell equations are newly derived.
\end{abstract}



\maketitle
\section{Introduction}
Ren\'{e} Descartes requested a system of science that explains both
mode and reason of phenomenon (i.e., how and why)\cite{edge}, but it
seems that he failed in explaining correctly either of them.
Subsequently, Sir Isaac Newton made a coup in explaining the mode of
phenomena, but even the Newtonian mechanics, which is the matrix of
current physics, failed in explaining the reason of phenomena. As
known in Newton's own endeavor\cite{Ladyman}, the correct
understanding of the reason of phenomena is indispensable for
completing a natural philosophy with consistency. Nevertheless, a
question about the reason of phenomena has been forgotten under the
admirable success of Newtonian mechanics that has been revealed in
description and prediction of phenomena.

This success of Newtonian physics results from adopting a
quantitative description of phenomena, which can be improved more
and more by comparing with experimental results. Here, the
quantitative description in theories of physics substantially
corresponds to mathematical abstraction, which is the major feature
of modern physics. Nonetheless, if an essence of mathematical
abstraction cannot be understood concretely, this abstraction leads
us to understand a phenomenon as just the phenomenon. In other
words, the mathematical abstraction is merely a quibble for evading
the essence of phenomenon and hinders us from understanding the
reason of phenomenon. In this sense, abstract concepts need to be
re-interpreted using substantial concepts in order that we have the
natural philosophy with consistency.

Meanwhile, some scientific philosophers have said that the reason of
phenomenon cannot be explained\cite{Nagel}\cite{EWHobson}. Of
course, if we have an interest in \textit{only describing an
empirical phenomenon by using abstract concepts}, it seems that
their despair is unavoidable. But, I believe that their despair can
be overcome. As will be shown later in this paper, \textit{careful
considerations to an entity and a process of cognizing it enable us
to understand the reason of phenomenon in a substantial level}.
These considerations will start from statements on the entity that
is a metaphysical subject. Nevertheless, it seems that our starting
statements, which will be given as postulates, can be highly
justified in several philosophical viewpoints and are compatible
with the known results of current physics. In addition to this
point, given the successful results that will be shown in this
paper, I think that questions related to phenomena, which have been
the subject of science up to now, can be reduced to metaphysical
questions as the subject of philosophy.

The construction of this paper is as follows. In Sec.II, I will
introduce some postulates to provide that an object of physical
inquiry (i.e., the matter and space), which will be called an
ex-entity, has transcendental, objective, independent, conservative
and singular characteristics. Here, it is worth noting that the
objective characteristic of ex-entity is the main basis of all
arguments that will be made in this paper. Thereafter, I will
introduce additional postulates that confine possible
existential-modes of the ex-entity; these postulates require that we
describe the magnitude, position and change of ex-entity. In
sequential consideration to the process of cognition, it will be
explained that the physical world can be recognized by only the
perception of dissimilarity. Next, we will discuss how to describe
the magnitude, position and change of ex-entity. In this discussion,
we will come to conclusions that the concept of density is required
for describing the magnitude of ex-entity and the density can be
written by a function of position and velocity.

In Sec.III.A, the mode, reason and magnitude of change will be
discussed on the basis of the objectivity of ex-entity. In
Sec.III.B, we will discuss methods of describing the magnitude,
position and change of ex-entity in order to establish a
precondition for describing objectively the physical world. In
Sec.III.C, we will obtain the law of motion, which prescribes a
relation between the motion of object and the external density of
ex-entity, on the basis of the objectivity and conservativeness of
ex-entity. Next, we will compare quantities of ex-entity that are
contained in the stationary cube and the moving cube, and from this
comparison, we will come to a conclusion that \textit{Lorentz
factor}, which is the keyword of special relativity,
\textit{represents a change of density caused by the movement of
object}. In Sec.III.D, we will discuss the origin of relativity,
which appears to be incompatible with the objectivity, on the basis
of the objectivity of ex-entity. From this discussion, we will see
that the theory of relativity can be explained from the objectivity
of ex-entity and the afore-mentioned objective description; that is,
we will verify that \textit{our conclusions related to the length
contraction, the time dilation and the Lorentz transformation
coincide with those of the theory of relativity}. In Sec.III.E, the
gravitational field will be considered in connection with the law of
motion obtained in Sec.III.C., and next, we will discuss how to
express mathematically the density distribution of ex-entity, which
generates the gravitational field. Especially, the fact that the
Lorentz factor represents the density of ex-entity will be
importantly used in this discussion. In this section, \textit{the
afore-mentioned reason of phenomenon will be answered
quantitatively}, and some issues related to the general theory of
relativity will be examined further.

The aim of Sec.III.F is to expand the idea suggested in this paper.
For this, we will discuss the electromagnetic and quantum mechanical
issues; e.g., a stability of matter, a force and field, a
relationship between electric charge, mass and quantity of
ex-entity, a spin, a size of particle, the Lorentz force and the
Maxwell equations. But, to tell the truth, I fail to develop
completely and sufficiently my arguments related to these issues,
because these issues are deeply connected with difficult problems
that have not solved in even the present physics. For all that, I
think that these issues merit reader's sober reflection--
particularly, \textit{the Lorentz force and Maxwell equations will
be plausibly derived from results obtained on the basis of the
objectivity and new acceptable assumptions such as a conservation of
momentum, in Sec.III.F.6.}
\section{Philosophical Consideration}
\subsection{Entity}
Let us define 'entity' as anything that can be said to exist and
'cognition system' as every mental process performed in the human
head. Then, the entity can be classified into 'ex-entity' and
'in-entity' depending on whether it exists inside or outside the
cognition system.
\subsubsection{Ex-entity}
According to this classification, the ex-entity corresponds to the
thing-in-itself mentioned by Kant and is also the subject of
physical science under the conviction of its objectivity. However,
since the ex-entity has a transcendental characteristic as will be
mentioned in the following postulate 1, it is the source of
consumptive arguments in that it can be interpreted in various ways
from philosophical viewpoints. In order to avert the consumptive
argument and make the starting premises of this paper clear, I will
introduce the following postulates concerning the ex-entity.
\begin{widetext}
\textsf{\textbf{\\
Postulate 1 : Ex-entity is a transcendental basis
that makes cognition possible.\\
Postulate 2 : Ex-entity exists objectively.\\
Postulate 3 : Ex-entity never disappears.\\
Postulate 4 : Ex-entity is of only one kind.\\
Postulate 5 : Ex-entity exists spatially.\\
Postulate 6 : Ex-entity changes.\\}}
\end{widetext}

Postulate 1 provides the transcendental characteristic of ex-entity
as a relationship between the ex-entity and the cognition. Postulate
2 represents that the ex-entity is an objective real existence that
is independent of the cognition system; therefore, we can say that
the ex-entity has objectivity and independence. Postulates 3 and 4
provide the conservative and singular characteristics of ex-entity
that are important for the following physical
consideration\footnote{The postulates 3 and 4 seem to be justified
from the postulates 1 and 2, but I will not discuss this subject in
this paper to avoid consumptive arguments.}\footnote{Given that the
ex-entity has objectivity by postulate 2, the postulate 3 is
distinguished from conservative characteristic of abstract concepts
such as the conservation of energy.}. In conclusion, from the
postulates 1 to 4, the ex-entity has the transcendental, objective,
independent, conservative and singular characteristics. Postulates 5
and 6 are statements on possible existential modes of ex-entity, as
will be discussed in detail later.

\subsubsection{In-entity}
The in-entity is the entity that exists within the cognition system,
constitutes the cognition system, and serves for cognitive
processes. If we exclude egregious mysticism, it is obvious that the
world outside the cognition system (i.e., the world of ex-entity or
the physical world) can be perceived by means of only sensory
perceptions on dissimilarities of ex-entity. (Hereinafter, we will
refer to the 'sensory perceptions' and the 'dissimilarities of
ex-entity' as "perceptions" and "dissimilarities", respectively, for
brevity's sake.) In other words, if there is no perceivable
dissimilarity, the cognition system is isolated from the external
world (i.e., the physical world). For this reason, it can be
concluded that every cognitive process starts from the perception of
dissimilarity, and most of abstract concepts, which do not represent
directly the dissimilarity of ex-entity, are obtained by processing
the perceived dissimilarity within the cognition system. For
example, energy is not measured directly by a sense organ or
measuring equipment; it is just a property of physical system that
is mentioned as something maintained constantly when the perceived
dissimilarities are processed in knowledge system of physics.

In the meantime, since the concept of energy is defined
mathematically, it is clear in the mathematical structure of physics
at least. But, the concept of energy is just a vague quibble in the
substantial aspect; that is, it is unclear what is constant. Of
course, such abstract concepts are manifestly useful to explain a
manner of phenomena (i.e., \textit{how}). Nevertheless, they are
substantially useless to study a reason of phenomena (i.e.,
\textit{why}), for their vagueness. Here, abstraction is a process
of generalizing phenomenal contents that are hard to be analyzed
using concrete concepts. In this sense, in order to overcome the
vagueness of abstract concepts, it is necessary to discuss a method
for representing the dissimilarity of ex-entity, which corresponds
to the most concrete and unartificial content. Only a description of
phenomena based on the dissimilarity enables us to analyze
substantially the quibbled concepts. For example, a substantial
analysis of energy will be seen in Sec.III.E.3.
\subsection{Dissimilarity}
From the postulates 5 and 6, the dissimilarity of ex-entity can be
classified into three independent components, namely, 'magnitude',
'position' and 'change', and \textit{only} such classification seems
to be valid in the light of physics. Hereinafter, dissimilarities in
magnitude, position and change will be referred to as $\Delta m$,
$\Delta p$ and $\Delta c$, respectively, for convenience.

\subsubsection{Minimum Magnitude for Distinction}
Let us define distinction as a cognitive process of perceiving the
dissimilarity of ex-entity, and let us define a minimum magnitude
for distinction $\Delta_{min}$ as a minimum magnitude of
dissimilarity that makes distinction possible. Here, $\Delta_{min}$
may decrease by improving sensitivity of measuring instrument.
Nevertheless, \emph{given that it is possible to cognize the
external world by only the perception of dissimilarity, it is
obvious that $\Delta_{min}$ cannot be zero}. For this reason,
$\Delta_{min}$ denotes an essential limitation of cognition, and the
cognition of $\Delta m$, $\Delta p$ and $\Delta c$ is subjected to
the corresponding $\Delta_{min}$.

\subsubsection{Magnitude}
The existence of an object can be recognized by perceiving its
$\Delta m$. Specifically, if there is no perceivable $\Delta m$
between an object and its vicinity, the existence of object cannot
be recognized. In addition, if no object can be recognized by
perceiving its $\Delta m$, either one of its $\Delta p$ and $\Delta
c$ cannot be perceived. It is necessary to remember this point to
understand a concept of complementarity, which will be discussed
later.

A degree of $\Delta m$ is typically described using a physical
concept of quantity. The concept of quantity can be understood as a
combination of the ex-entity for substance and the number for form.
Also, the concept of quantity seems to be most intuitive and
essential because it correlates directly the 'ex-entity' outside the
cognition system with the concept of 'be', which is the most
fundamental concept of the cognition system. Nevertheless, the
concept of quantity has arbitrariness because it does not have a
criterion for comparison; for example, a comparison of quantities
contained in two boxes having different volumes is generally
meaningless because of the volume difference. This arbitrariness in
the concept of quantity can be overcome by using the concept of
density that is defined as a quantity of ex-entity contained in unit
volume. (Unless there is any room for confusion, the 'density of
ex-entity' and the 'quantity of ex-entity' will now be referred to
as 'density' and 'quantity', respectively, for convenience.)

In the meantime, given the relation of mass and acceleration in the
Newtonian mechanics, the quantity or density of ex-entity cannot be
directly determined by perceiving the $\Delta m$ of ex-entity. The
density can be determined only through calculation using information
about $\Delta p$ and $\Delta c$. This density determination process
will be discussed in detail later.

\subsubsection{Position}
By perceiving $\Delta p$, we can recognize that one object is not
identical with the other. That is, unless $\Delta p$ between two
different objects can be perceived, we cannot know that the objects
are different from each other.

As is well known, a position and a degree of $\Delta p$ can be
described by using the spatial coordinates and spatial length,
respectively. Here, contrary to the density, the spatial length can
be determined through an observation. For example, in the case of
two meter-sticks, one's length can be directly compared with the
other's length by observing scales graduated on them, and this
comparison makes it possible to determine the spatial length. Of
course, quantities and densities of meter-sticks are determined by
means of not an observation but a calculation process based on
knowledge of physics, e.g., the Newton's second law.

In the meantime, as mentioned in the previous section, if there was
no object whose existence can be recognized, it would be impossible
for us to perceive $\Delta p$. In this sense, the perception of
$\Delta m$ is a precondition for perception of $\Delta p$. On the
contrary, if $\Delta m$ of objects are described without specifying
their positions, it is obscure which one of two objects is
described. For this reason, $\Delta m$ and $\Delta p$ are
complementary to each other.

Given this complementarity between $\Delta m$ and $\Delta p$, the
magnitude of the $\Delta m$ of ex-entity should always be described
in connection with the position of ex-entity, for the clarity of
representation. Accordingly, we will introduce a density
distribution function, $\rho =\rho(r)$, which expresses the density
of ex-entity as a function of position and is calculated from the
above-mentioned density determination process.

\subsection{Matter and Space}
According to the conventional physical viewpoint, it is understood
that matter is intrinsically of a different kind from space, and the
matter is subdivided into several fundamental particles according to
physical properties such as mass, electric charge and spin.
Furthermore, according to this viewpoint, the space is a vacuum
state, and the matter is an existing object that is wandering in the
space: it is generally taken for granted that there is a substantial
boundary surface between the matter and the space to separate being
and nothing.

But, from the singleness of postulate 4, the entity outside the
cognition system (i.e., ex-entity) is of only one kind. Thus, we can
say that the matter and space, which are definitely present outside
the cognition system, are not intrinsically of different kinds from
each other. Judging from the non-vanishment of minimum magnitude for
distinction and the significance of perception of $\Delta m$
explained in Sec.III.B, it can be explained that differentiation of
the matter from the space is based on not the kind of ex-entity but
the magnitude of density of ex-entity. That is, the matter
corresponds to a local region of ex-entity where existence can be
recognized by perceiving its $\Delta m$, and the space corresponds
to the other region of ex-entity where existence cannot be
recognized - where density is less than the minimum magnitude for
distinction in $\Delta m$.

From this analysis, the space is a non-empty portion of ex-entity.
Consequently, the length, which was introduced to express the degree
of $\Delta p$, can be understood as a physical magnitude that
represents the quantity of ex-entity corresponding to $\Delta p$:
the length can be described in terms of the quantity of ex-entity.
Specially, given that the quantity of ex-entity has essential
objectivity, which is the origin of physical objectivity,
\textit{the length must be described in terms of the quantity of
ex-entity for its objective description}. (The essential objectivity
of the quantity of ex-entity will be discussed in connection with
relativity in detail later.)

Although it was concluded that the space is not empty, this
conclusion should be distinguished from any attempt for resurrecting
the ether that was introduced to explain the propagation of light.
According to the ether hypotheses, the ether was interpreted as a
kind of matter that is different from conventional matters, such as
apples and electrons, and has transparent and undetectable
properties. Contrary to this, in our above conclusion, the matter
and space are regarded as entity of the same kind. In this sense,
our conclusion of space is definitely different from the ether
hypothesis. As is well known, the ether hypothesis in which the
ether is regarded as another matter is incompatible with the
Michelson-Moley experiment and the special relativity, but as we
shall see later, our conclusion of matter and space leads to results
that are compatible with them. Furthermore, our conclusion will
provide us with profound knowledge that has not been revealed in the
special theory of relativity.

Similarly, the matter cannot be classified on the basis of the kind
of ex-entity because of its singular characteristic; that is, the
matter is also of only one kind. In addition, given the
transcendental and independent characteristics of ex-entity, it can
be concluded that the physical properties for classifying the matter
do not exist objectively outside the cognition system. Rather, such
physical properties for classifying the matter are merely abstract
concepts that express phenomenal regularities detected by making
observations of the physical world. This is because the ex-entity
has only the a priori characteristics written in the above
postulates, but the regularities are one of a posteriori
characteristics that are learned only by experiences on ex-entity.
For this reason, it can be concluded that all the a posteriori
properties including the regularities originate from the a priori
characteristics of ex-entity. Particularly, the regularities result
from objectivity of ex-entity, as will be argued later.
\section{Physical Consideration}
The laws of physics are universal statements representing
regularities that can be learned from observations of physical
phenomena, and most of them are generally expressed by mathematical
equations, each of which prescribes a quantitative relation between
its left and right sides. In this aspect, physical regularity means
quantitative regularity that is found in a relation between physical
quantities. As we have discussed, the regularity itself should be
interpreted as the result of a priori characteristics (esp.,
objectivity) of ex-entity. Furthermore, in the following section
III.A, we will discuss the reason that the physical regularity can
be expressed in a quantitative form.
\subsection{Change}
Considering the postulate 1, knowledge of the world outside the
cognition system can be obtained from the experience of ex-entity.
Here, given the postulates 2 and 6, the change of ex-entity is an
objective actuality and enables us to experience the outer world. In
the following subsections, we will discuss the implications of the
aforementioned postulates in the mode, reason and magnitude of
change. Here, the following conclusions will be obtained from
explanations on the basis of the dissimilarity of ex-entity that can
be solely objectified; hence, they are statements having
cognition-independence. Therefore, the following conclusions should
be distinguished from both the phenomenal explanation of
Descartes\footnote{Descartes wanted to explain the movement of
matter on the basis of phenomena such as collision and vortex.
Although such phenomenal concepts have empirical intuitiveness, they
should be analyzed, in a substantial level, based on the
dissimilarity of ex-entity because phenomena are substantially only
the derivative results of dissimilarities of ex-entity as mentioned
above.} and the personified and teleological explanation such as the
least action principle\footnote{The least action principle demands
that a physical object should search for a course in which the
abstract quantity of action is minimized. But, it is obvious that
the ex-entity cannot do such physical thought.}.

\subsubsection{Mode of Change}
A mode of change of ex-entity is restricted by the postulates 3 and
4-conservative and singular characteristics. That is, since the
ex-entity never disappears and is of only one kind, it is hard to
escape a conclusion that the change of ex-entity is only the change
of density distribution. Especially, if we exclude mysterious
answers, this conclusion is inevitable. In this respect, we can
conclude that all phenomenal concepts, such as movement of particle,
wave and collision, are derived from processing of the perceived
change of density distribution: they are just derivative concepts.
In conclusion,\\
\\
\textsf{\textbf{[ Mode of Change ] - The change of ex-entity can be
only achieved by the change of density distribution.}}
\\

In this case, the change of ex-entity can be described in two ways -
a position-based description and a density-based description; the
former is the way of representing a change of density at a fixed
position, and the latter is the way of representing a change of
position having a fixed density. By comparison with the descriptive
ways of wave, the position-based description has affinity with a
method of expressing a change of amplitude of wave at a fixed
position, and the density-based description has affinity with a
method of expressing a change of position having fixed amplitude. Of
course, in both the descriptions, time is inevitably used as a
parameter to describe the change. (We will discuss the essence of
time in Sec.III.B.3-4.)

Notwithstanding, contrary to the vibration of string, the density of
ex-entity cannot be measured directly as discussed above. Hence, a
position-based description cannot be directly used to describe the
change of density distribution. In contrast to this, since we can
find an object with a fixed density in a restricted
scope\footnote{For example, if a change in $\Delta m$ cannot be
detected from a perceivable object such as matter and light, the
density of object can be interpreted as being constant at least
within the limit of minimum magnitude for distinction. Nevertheless,
the above-mentioned perceivability of object enables us to measure
the positional change (i.e., movement) of object. The density-based
description is therefore possible within this restriction.}, the
density-based description can be used to describe the change of
object: that is, we can describe effectively the movement of object
in the way of density-based description. For this reason, the
density-based description will be mainly adopted for the following
discussions related to the description of change, and we will use
some terms for the density-based description, such as matter,
movement, velocity and acceleration, if required.

However, it is worth noting that the density-based description can
be performed only on a perceivable object (e.g., matter and light):
it cannot provide us with any information on regions outside the
perceivable object. As a result, the density-based description can
be used to describe the motion of object but not to obtain the
density distribution of the entire space. In order to obtain the
density distribution of space, we need to establish a quantitative
relation between the density distribution and the movement of
object, as will be concretely discussed later.

\subsubsection{Reason of Change}
If we want to make physics objective, it is obvious that the reason
of change should be examined in connection with the ex-entity and
its three dissimilarities whose objectivity can be assured. Also,
explanations based on mysticism must be excluded in this
examination. These requirements related to the reason of change lead
to the conclusion that the dissimilarity of ex-entity causes a new
change (i.e., acceleration) of ex-entity. To put it more concretely,
it seems reasonable to conclude that the uniform motion of the
universe with a uniform density distribution can make no new change
in the moving state of an object that is co-moving with the
universe. We can therefore conclude that the non-uniform
distribution of ex-entity is the unique and general state of
dissimilarities of ex-entity that can change the moving state of the
object. Accordingly, the reason of acceleration can be concretely
expressed as follows:
\\ \\
\textsf{\textbf{[ Reason of Acceleration ] - A change in the moving
state of an object results from non-uniformity of density
distribution at a position where the object is located.}}
\\

This conclusion explains the reason of phenomena requested by
Descartes. In particular, since only the ex-entity and its
dissimilarities can be objectified as mentioned repeatedly above,
only the above conclusion is an explanation having
cognition-independence. Moreover, the above conclusion is the most
fundamental explanation on the origins of all physical changes in
that it is the universal statement without any confinement; it
implies that the physical regularity (i.e., the law of physics) is
singular in that the above conclusion restricts the reason of change
to only one.

Besides, according to the above conclusion, the concept of
action-at-a-distance must be excluded from physical considerations.
In spite of this exclusion of action-at-a-distance, in order to
explain interactions between distant objects, it is necessary to
introduce the concept of field, as is well known. But, the field
should be also related to the density distribution on the same
ground. For example, in the case of a gravitational force between
the earth and the sun, it can be understood that one object (e.g.,
the sun) is density distribution itself that can be expressed by a
certain density distribution function as mentioned above, and the
other object (e.g., the earth) is affected by the density
distribution of the sun. This is similar to the concept of scalar
potential except for the ontological reality of density; however, it
is obvious that the density distribution function is not identical
with the gravitational field in that the density distribution
function is a scalar field but the acceleration is a vector
quantity. A relation between the field and the density distribution
function will be minutely discussed later. Additionally, given the
sun's strong stability, it is highly probable that the density
distribution function of the sun has a particular mathematical
structure to maintain the sun as it is. This subject will be also
discussed later.

In addition to the new change of moving state (i.e., the
accelerative motion), the matter may move with a constant velocity
(i.e., uniform motion). From the above conclusion, we can see that
if the cause of change disappears (i.e., if a density distribution
become uniform), the moving state of an object is not changed. In
fact, this is identical with Newton's first/second laws.
Nonetheless, both the above conclusion and the Newton's laws are
explaining a condition for uniform motion, but not the essence of
motion\footnote{Note that the essence of motion means not a reason
of acceleration but that of movement itself.}; that is, we don't
know yet why movement occurs. We will make an answer to the essence
of motion on the basis of density and its conservativeness.

\subsubsection{Magnitude of Change}
If we accept the afore-mentioned reason of acceleration, it is
obvious that the resultant acceleration of object is only dependent
on the magnitude of density distribution in the neighborhood of the
object. Here, since the density distribution function
$\rho(\textbf{r})$, which was introduced in Sec.II.B.3, expresses
the distributional magnitude of density, the acceleration of object
should be represented in connection with the density distribution
function. Similarly, given that the uniform motion of matter is also
dependent on the density of space where the matter is located, we
can say that the velocity of matter is dependent on the density
distribution function. Beyond this qualitative dependence,
quantitative relations between the acceleration/velocity and the
density distribution function will be discussed in detail later. At
all events, the magnitude of change can be concluded as follows:
\\ \\
\textsf{\textbf{[ Magnitude of Change ] - Physical magnitudes
related to the moving state of object, e.g., velocity and
acceleration, are dependent on the density distribution function.}}
\\

In the meantime, given that the acceleration is dependent on the
density distribution function as stated in the above conclusion, we
can see that the density distribution function of space can be
determined in connection with the acceleration of phenomenon that
occurs at the very space. It is worth noting that this conclusion
enables us to justify the process of density determination, which
will be discussed later. In addition, since the density distribution
function, which determine the magnitude of change, can be
\textit{objectified} and \textit{quantified}, the above conclusion
explains not only the origins of physical quantitative regularities
but also the reason of every regularity related to the magnitude of
change, on the objective ground. Given that these quantitative
regularities are expressed by equations defining the quantitative
relations between physical magnitudes, not only descriptions of
respective physical magnitudes but also establishments of relations
between them should be objectified so that we can express the
quantitative regularity correctly.

\subsection{Objective Description of Physical Magnitude}
\subsubsection{Reference Magnitude and Ratio}
Every physical magnitude is represented as a ratio to a
predetermined reference magnitude. Concretely, a magnitude of
comparative object (hereinafter, a comparative magnitude) in a
certain physical substance is represented as a ratio to a magnitude
of predetermined reference object (hereinafter, a reference
magnitude) in the same physical substance. As a result, the physical
substance of comparative magnitude is expressed by the reference
magnitude, and a numerical relation between the reference and
comparative magnitudes is expressed by a ratio that is a
dimensionless number. Here, \textit{the reference magnitude is not
measured or calculated but merely defined as a unit value, and the
ratio is empirically determined by means of a measurement}. Hence,
in order to describe the physical magnitude objectively, both the
definition of reference magnitude and the determination of ratio
must be executed through objective methods.

Although a physical measurement has uncertainty that depends on the
minimum magnitude for distinction, its method is reliably objective.
That is, if two physicists measure ratios by using the same method
of measurement, there is no denying the objectivity of measured
ratios. Of course, a wrong measurement gives rise to a wrong result,
but this is irrelevant to the topic of objectivity under discussion.

Given that the reference magnitude is just defined as the unity,
even if two reference magnitudes defined independently by two
physicists are expressed by the same number (i.e., unity), they may
be substantially different from each other. For example, both one
meter and one inch are identically expressed by the number of one,
but there is a manifest spatial difference between them. The units
of 'meter' and 'inch' are used to differentiate such substantial
difference. In conclusion, one reference magnitude defined by one
physicist is objective for the physicist's own sake but not for the
others, and two reference magnitudes defined independently by two
physicists cannot be objectified until a substantial difference
between them is revealed (i.e., converted) quantitatively.

In the meantime, given the objectivity of ex-entity, it is obvious
that an object for defining the reference magnitude can be freely
selected, and that reference magnitudes can be objectively converted
to each other. Nevertheless, it is necessary to take facility of
conversion between reference magnitudes into consideration, because
the conversion is actually one of complex procedures for determining
a ratio between magnitudes. In particular, the dissimilarities of
ex-entity are the most fundamental components that represent the
possible existential mode of ex-entity as discussed above, thus we
will discuss how to define the reference magnitudes of
dissimilarities with regard to the facility of conversion, in the
following section.
\subsubsection{Reference Magnitudes of Density and Length}
As discussed in Sec.II.B.2-3, the density and length are physical
quantities that represent $\Delta m$ and $\Delta p$, respectively.
Given that an object for defining the reference magnitude can be
freely selected as mentioned in the previous section, space can be
selected as a reference object for defining reference magnitudes of
density and length. Especially, this selection of space is justified
from the fact that the space is a portion of ex-entity with
objectivity as discussed in Sec.II.C. (Hereinafter, we will refer to
the reference magnitude of density as 'reference density' or 'unit
density' and the reference magnitude of length as 'reference length'
or 'unit length', for brevity's sake.) For example, an observer A
can select an arbitrary position $\mathbf{r}_{A}$ in space to define
his reference density $\rho^{A}_{0}$, [i.e.,
$\rho^{A}_{0}กี\rho(\mathbf{r}_{A})$], and select a distance between
two arbitrary positions $\mathbf{r}_{1}$ and $\textbf{r}_{2}$ to
define his reference length $L^{A}_{0}$, (i.e.,
$L^{A}_{0}กี|\mathbf{r}_1 -\mathbf{r}_2|$).

This reference density $\rho^{A}_{0}$ serves as a standard for
describing the density distribution function $\rho(\mathbf{r})$,
which was introduced in Sec.II.B.3. For this description, let us
introduce the distribution factor $\phi(\mathbf{r})$ that represents
a ratio of a density at a position \textbf{r} to the reference
density $\rho^{A}_{0}$ - a spatial variation of density. Then, the
density distribution function described by the observer A can be
given by
\begin{equation}
  \rho(\mathbf{r})=\rho^A_0
  \phi(\mathbf{r})=\rho(\mathbf{r}_A)\phi(\mathbf{r}).
\end{equation}
From the definition of density, $\rho(\mathbf{r})$ of Eq. (1)
represents a quantity of ex-entity contained in the unit volume at
\textbf{r}. This meaning of $\rho(\mathbf{r})$ should be remembered,
because it is related to the problem of singularity as will be
discussed in the Sec.III.F.4.

In the meantime, similar to the case of the observer A, another
observer B can select other position $\mathbf{r}_B$ to define his
reference density $\rho^{B}_{0}$, [i.e., $\rho^{B}_{0}\equiv
\rho(\mathbf{r}_B)$]. But, as mentioned above, the reference
magnitudes are defined as merely unity and may make a substantial
difference. That is, even if $\rho^{A}_{0}$ and $\rho^{B}_{0}$ are
expressed by the same value of unity, they may different from each
other, because of the positional difference between $\mathbf{r}_A$
and $\mathbf{r}_B$. The difference between $\rho^{A}_{0}$ and
$\rho^{B}_{0}$ can be written by
\begin{equation}
  \rho^B_0=\rho^A_0\phi(\mathbf{r}_B).
\end{equation}
Nevertheless, it is not until the distribution factor (eventually,
the density distribution function) is determined that we can know
the substantial difference.

As discussed in Sec.II.C, the length, which was introduced to
represent a degree of $\Delta p$, expresses the quantity of
ex-entity corresponding to $\Delta p$. In order to express a
relation between length and its corresponding quantity, let us
define length quantity $Q_L$ as the quantity of ex-entity
corresponding to a length L. Then, from the definition of density,
the quantitative relation between L and $Q_L$ is given by
\begin{equation}
  L=Q_L / A\rho,
\end{equation}
where $\rho$ denotes a density of space where the length L is
measured, and the term A denotes a unit area perpendicular to the
direction of L so that a volume equation of V=AL is satisfied. As a
result, we can conclude that the relation between L and $Q_L$ is
also dependent on the density. Hence, the reference volumes and the
reference lengths that are respectively defined by the observers A
and B may make a substantial difference depending on the densities
of spaces where they are defined. Similar to the above argument on
density, substantial differences in volume and length cannot be
revealed quantitatively until a function of density is known
quantitatively.
\subsubsection{Change and Time}
As is well known, time is used as a standard for describing the
magnitude of change (hereinafter, 'a reference magnitude of
change'). It seems that this usage of time as the reference
magnitude of change results from a peculiar property of time that
can be said as uniform passage. In that case, is time an ontological
real existence having the property of uniform passage? Or, do
physical phenomena keep in step with the passage of time?

Time cannot be understood as a real existence having ontological
objectivity, because the ex-entity and its dissimilarities are the
only ontologically objective things as discussed above. Especially,
if we exclude the personification of the physical world, it is
obviously impossible that physical phenomena keep voluntarily in
step with the passage of time, even though such personified
interpretation is greatly useful for physical descriptions. Then,
what is the uniform passage of time? It seems that this is related
to the regularity of magnitudes of changes that can be perceived
from various phenomena.

To put it more concretely, let us suppose that change-magnitudes of
phenomena P1, P2 and P3 remains the constant ratios of l:m:n. This
relation holds approximately true for the most cases comprising
movements of pendulum, sun, moon, earth, and light. Here, the
change-magnitude of one of them can be selected as a reference to
describe those of the others. For example, the change-magnitudes of
P2 and P3 can be expressed in constant ratios to P1 (i.e., m/l and
n/l). It seems that our clocks have been fabricated on the basis of
this relation among change-magnitudes of phenomena. Of course, this
constant ratio relation does not hold true for movements of
free-falling apple and accelerating car. However, magnitudes of such
accelerative movements are also not perfectly random in that they
have quantitative regularities that can be described by the
change-magnitude of P1.

In this sense, the uniformity of passage means just the constancy in
ratios of change-magnitudes (hereinafter, constancy in change
ratio). Here, it is obvious that the constancy in change ratio
results from the regularity of change-magnitude mentioned in
Sec.III.A.3. As discussed there, the regularity of change-magnitude
results from the fact that every change-magnitude depends on the
density of ex-entity that can be objectified. Consequently, we can
say that the afore-mentioned peculiar property of time (i.e., the
constancy in change ratio or the uniformity of passage) results from
the objectivity of ex-entity and its dissimilarities. As a result,
the magnitude of time should be also expressed in connection with
the density of ex-entity.
\subsubsection{Reference Magnitude of Change}
Given the freedom of selecting a reference, an arbitrary phenomenon
can be selected as \textit{a phenomenon for defining the reference
magnitude of change }(hereinafter, reference phenomenon).
Furthermore, as discussed in Sec.III.A.1, the change of ex-entity
can be only accomplished by the change of density distribution, and
this change of density distribution can be validly expressed by the
density-based description that represents a change of position
having a fixed density. Considering these conclusions, it is clear
that \textit{the reference magnitude of change can be defined by an
advancing length of reference phenomenon}. (Here, the advancing
length of reference phenomenon means a magnitude in positional
change of point having a fixed density. For brevity's sake, we will
now refer to the advancing length of reference phenomenon as a
reference advancing length.) In conclusion, \textit{the reference
magnitude of change can be represented by using a spatial length}.

Furthermore, since the spatial length represents the corresponding
quantity of ex-entity as mentioned above, \textit{the reference
magnitude of change can be represented by using a quantity of
ex-entity}. Similar to the Eq. (3), the reference advancing length
$L_T$ can therefore be expressed in terms of a corresponding
quantity of ex-entity $Q_T$ (hereinafter, time quantity) as follows:
\begin{equation}
  L_T = Q_T/A\rho
\end{equation}
where $\rho$ denotes the density of space where the reference
phenomenon takes place, and the term A denotes the unit area of
reference phenomenon perpendicular to the advancing direction of
reference phenomenon.

The reference advancing length $L_T$ defined by this way can be used
for two purposes -- a \textit{common reference} for describing the
change-magnitude of every phenomenon in a system and a
\textit{specific reference} for describing that of individual
phenomenon. The $L_T$ as the common reference serves as a parameter
for describing diverse phenomena and, eventually, corresponds to the
parametric time that is generally used for our physical
descriptions. That is, \textit{an observer can determine the
magnitude of temporal passage (i.e., a temporal length) in his
system by measuring the $L_T$}. On the contrary, \textit{the $L_T$
as the specific reference is used to describe speeds of individual
phenomena}, as will be discussed in the next paragraph. Of course,
the speed can be also described by the parametric time, as usual;
that is, the parametric time can take the place of $L_T$ that is
used as the specific reference. In fact, this conclusion is natural
in that the parametric time -- the reference magnitude of change --
can be expressed by a spatial length.

Let us discuss further a description of speed using the reference
advancing length $L_T$. Since the change is literally not static
unlike the length or the density, a phenomenon having a fixed
change-magnitude cannot be used as an objective reference for
describing diverse phenomena any more. A change-magnitude of each
phenomenon should therefore be expressed by a ratio to that of a
covariant reference phenomenon as follows:
\begin{equation}
  \frac{[advancing~length~of~comparative~phenomenon]}
       {[advancing~length~of~reference~phenomenon,L_T]}\nonumber.
\end{equation}

Given that a standard for comparison (i.e., $L_T$) is clearly
specified in the above definition, the magnitude of change described
in this way is objective. In addition, given that the reference
advancing length serves as the parametric time, it is obvious that
the change-magnitude expressed by this way is equivalent to the
aforementioned speed of each phenomenon. Particularly, we can say
that the speed is substantially a dimensionless
magnitude\footnote{The fundamental quantities and mathematical
structure of physics need to be further discussed in connection with
the conclusions that 1) the time can be expressed by spatial length
and 2) the speed is substantially dimensionless. These subjects will
however be not discussed anymore, because they have no relevance to
the aim of the present article.}, because both the numerator and
denominator of the above expression have dimensions of spatial
length. In addition, we can say that the reference magnitude, which
is the basis of objectification, is already implied into the
definition of speed: the speed itself is a physical concept having
its reference magnitude. In this sense, the length and the density
are distinguished from the speed, because they are objectified only
when their reference magnitudes are specified.

Meanwhile, the Lorentz factor is expressed by a ratio of speeds of
object and light (i.e., v/c), but in this ratio term, the dimension
of time in the numerator and denominator are canceled each other. We
can therefore say that the Lorentz factor is actually a function
depending on a ratio of an advancing length of object to that of
light. Furthermore, if the advancing length of light satisfies some
requisites for the reference advancing length, we can say that this
ratio term is also equivalent to the afore-defined speed. Of course,
given the freedom of selecting reference, it is natural that the
advancing length of light can be used as the reference advancing
length, and moreover, the light can be preferred as the reference
phenomenon, for its peculiarity. In the following Sec.III.C.2, we
will concretely discuss this peculiarity of light in connection with
features of space and light; for instance, the facts that 1) the
space is the common ground where every phenomenon occurs, and 2) the
speed of light is entirely dependent on the density of space.

Let us discuss the density dependence of temporal length. Similar to
the case of spatial length, since $L_T$ is dependent on the density
as written in Eq. (4), $L_T$ corresponding to the same $Q_T$ is also
changed with density. That is, Eq. (4) implies the relativistic
conclusion that the time is not a physical quantity regardless of a
state of system. In the Sec.III.D.3, we will see that the famous
relativistic time dilation can be explained from this density
dependence of temporal length. Meanwhile, the spatial and temporal
lengths can be mathematically expressed by a \textit{speed} of
object and a \textit{position} in a gravitational field, as shown
respectively in the Lorentz transformation and the general
relativity. We can therefore say that the spatial and temporal
lengths have physical regularities. Here, considering the origin of
quantitative regularity discussed in the Sec.III.A.3, we can
conclude that these regularities of spatial and temporal lengths
result from the objectivity of ex-entity (especially, its density).
The equations (3) and (4) are the quantitative explanation that
reconfirms this conclusion.

Finally, as we have seen, a phenomenon having a fixed
change-magnitude cannot be selected as the reference phenomenon: the
reference magnitude of change itself -- time -- changes ceaselessly
unlike the reference magnitude of length. For this reason, if $L_T$
is used as the common reference for expressing other change
phenomena (i.e., as time), we need to define additionally its
reference magnitude (i.e., a unit time) for describing the magnitude
of $L_T$ objectively. Of course, the unit time must be also defined
in connection with the quantity of ex-entity. For example, the unit
time can be defined as a time needed for the reference phenomenon to
advance the reference length. In this case, \textit{even if the
density of space varies, we can see that the unit time and the unit
length of system are changed at the same rate regardless of the
density of space}. In addition, given the objectivity of ex-entity
and the quantitative regularity derived from it, it is obvious that
\textit{the times needed for same phenomena (i.e., phenomena that
are subject to the same mechanism) to advance the same quantity of
ex-entity are equal to each other}. I believe that these two
conclusions, which are originated from the objectivity of ex-entity,
enable us to explain the first postulate of relativity, i.e., the
constancy of light speed. That is, the constancy of light speed is
just other statement that expresses the above conclusions based on
the objectivity.

\subsection{Density Determination I}
From the preceding arguments, the density is the physical magnitude
that represents the substance of ex-entity (i.e., the existence or
the $\Delta m$), and the spatial coordinates and the velocity are
magnitudes for representing the existential modes of ex-entity
(i.e., the spatial dissimilarity and the change). And, from the
postulate 2, 5 and 6, these three kinds of dissimilarities can be
only objectified. We can therefore conclude that \textit{every
difference in a physical substance (or content) must be capable of
being completely expressed by a density function of ex-entity}, and
\textit{this density function must be capable of being described by
using the spatial coordinates and velocity as variables}. For these
reasons, in addition to the distribution factor that describes a
change of density caused by a positional difference, we need to
introduce a kinetic factor, which is a function of velocity (i.e.,
that of change magnitude), to describe a change of density caused by
a movement.
\subsubsection{The Law of Motion}
From the discussion in the Sec.III.A.1, a non-uniform density
distribution causes a change of movement state of object. In this
section, we will concretely discuss a quantitative relation between
the distribution factor and the kinetic factor on the basis of
objectivity of ex-entity. As written in Eq. (1), in this case, the
observer A will describe a density distribution of space as follows:
\[\rho(\mathbf{r})=\rho^A_0\phi(\mathbf{r})=\rho(\mathbf{r}_A)\phi(\mathbf{r}).\]
Here, $\rho^A_0$ -- the reference density for the observer A --
denotes the density of ex-entity at the reference position
$\mathbf{r}_A$, which is selected by the observer A, and is just
defined as unity in value. Contrary to this, the distribution factor
$\phi(\mathbf{r})$ is an unknown function because $\Delta m$ cannot
be measured directly as mentioned above. Meanwhile, since the
reference density is dependent on a position selected by an
observer, the reference density $\rho^A_0$ may vary with the
reference position $\mathbf{r}_A$. Hence, let us assume for
convenience that the reference position $\mathbf{r}_A$ is fixed with
respect to the observer A. Then, $\rho^A_0$ is a time-independent
constant.

Now, let us denote the density, position and velocity of test
object, which are described by the observer A, as $\rho^A_p$,
$\mathbf{r}^A_p$ and $\mathbf{v}^A_p$, respectively. Here, a ratio
of $\rho^A_p$ to $\rho^A_0$ may vary with $\mathbf{r}^A_p$ and
$\mathbf{v}^A_p$, but at this stage, we cannot know the ratio owing
to the impossibility of measuring the $\Delta m$. We can therefore
express $\rho^A_p$ by using an unknown ratio X to $\rho^A_0$; that
is, $\rho^A_p=\rho^A_0 X(\mathbf{r}^A_p, \mathbf{v}^A_p)$. And, the
position $\mathbf{r}^A_p$ varies with the movement of test object,
thus it can be described by using time as a parameter:
$\mathbf{r}^A_p=\mathbf{r}^A_p(t)$. Since the velocity
$\mathbf{v}^A_p$ denotes the velocity of test object at
$\mathbf{r}^A_p$, it is dependent on $\mathbf{r}^A_p$:
$\mathbf{v}^A_p=\mathbf{v}^A_p(\mathbf{r}^A_p)$. As a result, the
density of test object described by the observer A- $\rho^A_p$ - can
be written by
\begin{equation}
  \rho^A_p(t)=\rho^A_0~X\{\mathbf{r}^A_p(t),\mathbf{v}^A_p[\mathbf{r}^A_p(t)]\}.
\end{equation}

In the meantime, the test object can be described in the same manner
by another observer B who is co-moving with the test object. That
is, let us assume that a reference density for B (i.e., $\rho^B_0$)
is defined by the density of ex-entity at the position
$\mathbf{r}_B$, which is fixed with respect to the observer B. Then,
similar to the observer A, the observer B will describe the density
of test object as follows:
\begin{equation}
  \rho^B_p(t)=\rho^B_0~Y\{\mathbf{r}^B_p(t),\mathbf{v}^B_p[\mathbf{r}^B_p(t)]\},
\end{equation}
where $\rho^B_p$, $\mathbf{r}^B_p$ and $\mathbf{v}^B_p$ denote the
density, position and velocity of test object, which are described
by the observer B, and Y denotes an unknown ratio of $\rho^B_p$ to
$\rho^B_0$ for the observer B. Here, similar to $\rho^A_0$,
$\rho^B_0(t)$ is a density at the position that is fixed with
respect to the co-moving observer B, thus it is also a
time-independent constant to the observer B: $\rho^B_0(t)$ is a
constant that is unity in value. In addition, since the observer B
is co-moving with the test object, we have
\begin{subequations}
\begin{eqnarray}
  \mathbf{r}^B_p(t)&=&\mathbf{r}^B_p(0)\nonumber\\
  &=&\mathbf{r}^A_p(t)-\mathbf{r}^A_B(t)~~~~~~~~~~~~~(const.),\\
  \mathbf{v}^B_p(t)&=&\mathbf{v}^B_p(0)\nonumber\\
  &=&\mathbf{v}^A_p[\mathbf{r}^A_p(t)]-\mathbf{v}^A_B[\mathbf{r}^A_B(t)]\nonumber\\
  &=&\mathbf{0}~~~~~~~~~~(const.),
\end{eqnarray}
\end{subequations}
where $\mathbf{r}^A_B$ and $\mathbf{v}^A_B$ denote the position and
velocity of the observer B, which are described by the observer A.

Similar to the conventional physical consideration, let us assume
for convenience that the movement of test object is a complete
kinematical phenomenon that is not accompanied by a thermal or
internal process\footnote{For an imperfect kinematical process, it
seems that a careful consideration is needed. But we will not
discuss this subject here.}. Then, although the ratio Y is still an
unknown number for B, the ratio Y is independent of the movement of
test object, contrary to the ratio X for the observer A; that is, Y
is constant regardless of the movement of test object. As a result,
$\rho^B_p(t)$ - the density of test object written by the co-moving
observer B - is always a time-independent constant. That is,
\begin{equation}
  \rho^B_p(t)=\rho^B_0 ~Y~~(const.).
\end{equation}

But, considering that the reference position $\mathbf{r}_B$ where
$\rho^B_0(t)$ is defined is moving with respect to the fixed
observer A, the reference density $\rho^B_0(t)$ is not constant for
the observer A. As discussed in Eq. (2), this variation of
$\rho^B_0(t)$ can be revealed when $\rho^B_0(t)$ is described on the
basis of $\rho^A_0$ -- the reference density for A. That is, a
substantial magnitude of $\rho^B_0(t)$ can be given by the product
of the reference density for A and a magnitude of distribution
factor at $\mathbf{r}^A_B(t)$ as follows:
\begin{equation}
  \rho^{~A}_{0B}(t)=\rho^A_0 ~\phi(\mathbf{r}^A_B(t)),
\end{equation}
where $\rho^{~A}_{0B}(t)$ denotes the magnitude of
$\rho^{B}_{0}(t)$, at t=t, that is converted in terms of $\rho^A_0$;
that is, $\rho^{~A}_{0B}(t)$ is the magnitude of $\rho^{B}_{0}(t)$
that is described by A. Using Eq. (9), we can express
$\rho^{B}_{p}(t)$ of Eq. (8) in terms of $\rho^A_0$. That is,
substituting $\rho^{~A}_{0B}(t)$ of Eq. (9) into $\rho^B_0$ of Eq.
(8), we have
\begin{equation}
  \rho^{~A}_{pB}(t)=\rho^A_0 ~\phi(\mathbf{r}^A_B(t))~Y,
\end{equation}
where $\rho^{~A}_{pB}(t)$ denotes the magnitude of $\rho^{B}_{p}(t)$
that is converted in terms of $\rho^A_0$; that is,
$\rho^{~A}_{pB}(t)$ is the magnitude of $\rho^{B}_{p}(t)$ that is
described by A.

Now, let us discuss a relation between the unknown ratios X and Y.
For this, let us assume that at t=0, both the test object and the
observer B are at rest relative to the observer A. Then, from Eq.
(5), the initial density of test object described by A is given by
\begin{equation}
  \rho^{A}_{p}(0)=\rho^A_0 ~X(\mathbf{r}^A_p(0),\mathbf{0}).
\end{equation}
The initial density of test object can be similarly described by the
observer B. But, for the objective comparison, we need to convert
the initial density of test object described by B in terms of
$\rho^A_0$. That is, from Eq. (10), we have
\begin{equation}
  \rho^{~A}_{pB}(0)=\rho^A_0 ~\phi(\mathbf{r}^A_B(0))~Y.
\end{equation}
Given the objectivity of ex-entity density, Eqs. (11) and (12) must
be equal to each other, because they express the identical substance
(i.e., the initial density of test object) using the common
reference density (i.e., $\rho^A_0$). Consequently, we have
\begin{equation}
  X(\mathbf{r}^A_p(0),\mathbf{0})=\phi(\mathbf{r}^A_B(0))~Y.
\end{equation}

In the meantime, since the ratio X of Eq. (5) is not known, all the
observer A can say at t=t is only the fact that the test object
whose initial density factor was $X[\mathbf{r}_p^A(0), \mathbf{0}]$
is moving with a velocity $\mathbf{v}^A_p[\mathbf{r}^A_p(t)]$ at a
position $\mathbf{r}^A_p(t)$ at t=t. This statement of A can be
mathematically written using the kinetic factor as follows:
\begin{equation}
  \rho^A_p(t)=\rho^A_0 ~X[\mathbf{r}^A_p(0),
        \mathbf{0}]~\gamma\{\mathbf{v}^A_p[\mathbf{r}^A_p(t)]\}.
\end{equation}
Similarly, since Eqs. (10) and (14) also express the density of same
test object at the same time using the common reference density
(i.e., $\rho^A_0$), they must be equal to each other due to the
objectivity of ex-entity density. That is, we have
\begin{equation}
\phi[\mathbf{r}^A_B(t)]Y= X[\mathbf{r}^A_p(0),
        \mathbf{0}]~\gamma\{\mathbf{v}^A_p[\mathbf{r}^A_p(t)]\}.
\end{equation}
Substituting Eqs. (7) and (13) into Eq. (15), we have
\begin{equation}
 \frac{\gamma\{\mathbf{v}^A_B[\mathbf{r}^A_B(t)]\}}{\phi\{\mathbf{r}^A_B(t)\}}
        =\frac{1}{\phi\{\mathbf{r}^A_B(0)\}}.
\end{equation}

The above equation has only the position and velocity of the
observer B that are described by the observer A \textit{as
variables}. Hence, let us remove superscriptions and subscriptions
from the above equation, for convenience. In addition, although the
left side of the above equation is a function of time that expresses
the ratio of the kinetic factor to the distribution factor, it is a
time-independent constant because its right side is constant. Hence,
if we define the left side of the above equation as a ratio function
M, the above equation can be written in the simple form as follows:
\begin{eqnarray}
 M[\mathbf{r}(t), \mathbf{v}(t)]&\equiv&
 \frac{\gamma[\mathbf{v}(t)]}{\phi[\mathbf{r}(t)]}\nonumber\\
 &=&\frac{1}{\phi[\mathbf{r}(0)]}~~~(const.).
\end{eqnarray}
This equation is a universal statement obtained from the objectivity
of ex-entity density and prescribes the mode of movement that should
be obeyed by the test object. In this sense, I will designate Eq.
(17) as the law of motion of ex-entity hereinafter. Here, since the
ratio function M included in the law of motion is the
time-independent constant as mentioned above, it can be understood
as the constant of motion that plays an important role in a physical
analysis. We will see in Sec.III.E.3 that the energy, which is the
famous constant of motion, is closely related to the ratio function
M.

Meanwhile, if the reference positions $\mathbf{r}_A$ and
$\mathbf{r}_B$ coincide with each other at t=0, the ratio function M
becomes unity by Eq. (17); that is, the kinetic factor is always
equal to the distribution factor. Here, it can be comprehended that
the kinetic factor represents the internal density of test object,
while the distribution factor represents the density of space that
causes the movement of test object. In this sense, we can conclude
that a test object moves such that its internal density is equal to
the density of external space. In order to make a profound
comprehension of this conclusion, we will now discuss a relation
between speed and density and possible modes of change of density.

\subsubsection{Determination of Kinetic Factor}
In order to establish a quantitative relation between the kinetic
factor and the velocity, let us make a comparison of quantities of
ex-entity contained in two virtual cubes having the same volume.
First, let us assume that one cube is at rest relative to an
observer A and the other cube is moving with velocity \textbf{v}
relative to the observer A. (See FIGs. 1 and 2.) In addition, we
will assume that each of the cubes has a fixed volume regardless of
its movement. Of course, this assumption is incompatible with the
theory of relativity and the Michelson-Moley experiment. But this
assumption is just suggested to \textit{objectively compare
quantities of ex-entity contained in the virtual cubes}. That is, in
the following Sec.III.D, we will come to the relativity-compatible
conclusion that \textit{the volume of a real cube must be contracted
in its moving direction}.

Before making the comparison of quantities in earnest, let us
discuss how to determine a quantity of ex-entity contained in the
virtual cube. The quantity of ex-entity in the cube can be
determined by using the relation between the density and the
advancing length of reference phenomenon, as written in Eq. (4). A
matter may however move with various speeds depending on its
physical circumstance. Hence, if the physical circumstance is not
specified, it is hard to select a movement of matter as the
reference phenomenon for describing the changes of ex-entity.
Contrary to the matter, the speed of light is dependent only on the
properties of space (e.g., permittivity and permeability), thus the
light can be desirably selected as the reference
phenomenon\footnote{Considering that every physical substance can be
completely expressed by using the density function as discussed
above, the permittivity and permeability should be understood as
physical quantities that depend on the density. This subject will be
further discussed in Sec. III.F.6 related to the electromagnetism.
Meanwhile, if the physical condition is exactly known, it is obvious
that the speed of matter can also be used for the reference
magnitude of change and moreover, converted into the speed of light.
}. Especially, given that the space is the common ground where every
physical phenomenon occurs, the speed of light, which depends only
on the properties of space, is enough to be the reference magnitude
of change for describing change magnitudes of various phenomena. For
this reason, we will use the speed of light as the reference
magnitude of change to determine a quantity of ex-entity contained
in the virtual cube.

From Eq. (4), the advancing length of reference phenomenon
represents the corresponding quantity of ex-entity. Hence, by
measuring a time taken for the light to travel from one surface of
cube to the opposite surface thereof, we can compare the quantities
of ex-entity contained in the stationary and moving cubes. Here,
note that the compared volumes of cubes should be equal in two
cases\footnote{This requirement should be importantly considered
when the advancing direction of measurement light is parallel to the
moving direction of moving cube.} to make the comparison exact.

First, let us consider the case of rest cube. For convenience, let
us assume that the advancing direction of measurement light is
perpendicular to the moving direction of moving cube. That is, we
can select the regular tetragon defined by four points A, B, C and D
as a starting surface from which the measurement light starts, as
shown in FIG. 1. Since the cube under consideration is at rest, the
time $t_s$, which is taken for the light to travel from the points
A, B, C and D to the points E, F, G and H, is given by
\begin{equation}
 \mathrm{t_{s}}=\frac{l_{0}}{c},
\end{equation}
where $l_0$ denotes the length of each side of the cube and c
denotes the speed of light.
\begin{figure}
\epsfig{file=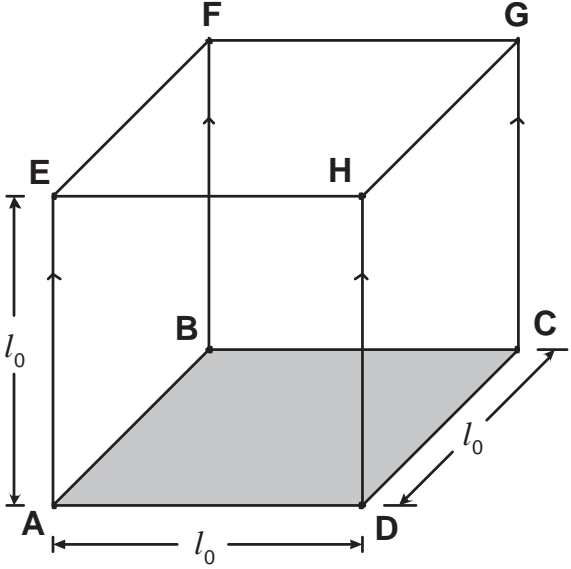, width=0.425\textwidth} \caption{Stationary
Virtual Cube : Volume=$l_0^3$.\label{FIG1}}
\end{figure}
\begin{figure}
\epsfig{file=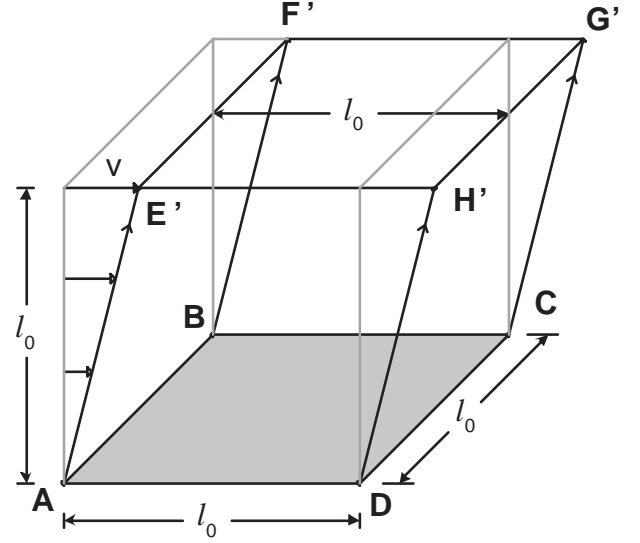, width=0.455\textwidth} \caption{Moving
Virtual Cube : Volume=$l_0^3$.\label{FIG2}}
\end{figure}

Now, let us consider the case of moving cube. Just as the case of
rest cube, let us select the regular tetragon defined by points A,
B, C and D as the starting surface. But, since this cube is moving
with a velocity \textbf{v} relative to the observer A in the
direction parallel to the starting surface, the observer A will
observe that the light beams, which start from the points A, B, C
and D, arrive at points E$'$, F$'$, G$'$ and H$'$, which are shifted
from the original arrival points E, F, G and H, respectively. See
FIG. 2. In this case, the time $t_m$ measured in the moving cube can
be calculated in the same way as in the conventional relativistic
argument on the time dilation\cite{Gasiorowicz1}. That is, by the
Pythagorean theorem, $t_m$ is given by
\begin{eqnarray}
 \mathrm{t_{m}}&=&\frac{l_{0}}{c}\frac{1}{\sqrt{1-\mathrm{v}^2/c^2}}\nonumber\\
                &=&\mathrm{t_{s}}\frac{1}{\sqrt{1-\mathrm{v}^2/c^2}}.
\end{eqnarray}

Here, note that the stationary and moving cubes have not only the
same area of starting surface but also the same volume. As mentioned
above, $t_s$ and $t_m$ represent the quantities of ex-entity
contained in the stationary and moving cubes, respectively. We can
therefore conclude that, from a comparison between Eqs. (18) and
(19), the density of cube moving with the velocity \textbf{v} is
equal to the product of the density of stationary cube and the
Lorentz factor as follows:
\begin{equation}
 \rho'=\rho \frac{1}{\sqrt{1-\mathrm{v}^2/c^2}}.
\end{equation}
where $\rho'$ and $\rho$ denote the densities of stationary and
moving cubes, respectively. From this result, we can say that the
Lorentz factor, which is a keyword of special relativity, is the
kinetic factor that expresses the change of density induced by a
movement of object. That is,
\begin{equation}
 \gamma(\mathbf{v})=\frac{1}{\sqrt{1-\mathrm{v}^2/c^2}}.
\end{equation}

Meanwhile, in the above thought experiment, the virtual cube was
introduced to mark merely the boundary of fixed volume. In this
sense, the ex-entity quantity calculated in the above argument
corresponds to the ex-entity quantity contained in the local space
that is defined by the virtual cubes. That is, the quantity
considered in the above argument is the quantity of space confined
by the cube rather than the cube's own quantity. Nevertheless,
considering Eq. (17), we can say that an object's own density does
also increase in the ratio of Eq. (21) with the object's velocity.
\subsection{Relativistic Consideration}
\subsubsection{Relativity}
Let us apply the same thought experiment to the case of a new
observer who is co-moving with the moving cube. Then, the new
observer will come to the same conclusion that the previous observer
A has obtained, similar to the special relativity. That is, the new
observer will conclude that the moving cube, which was at rest
relative to A, has an increased density more than the rest cube that
was moving relative to A. Nonetheless, since such relative
description between two observers seems to be contradictory to the
fact that the density is an objective magnitude, we should explain
the reason why such relative description is possible. For this, it
is necessary to discriminate between an \textit{essential
objectivity} and a \textit{descriptive objectivity}.

Given the objectivity of ex-entity, the quantity of ex-entity is
essentially objective; therefore, a relative description of
ex-entity quantity is meaningless and is not allowed to assure the
physical regularity. Contrary to this, it can be stated that the
length and time have only descriptive objectivity because they are
merely magnitudes that can be objectively described using the
quantity of ex-entity, as shown in Eqs. (3) and (4). Of course, from
the postulates 5 and 6, it is obvious that $\Delta p$ and $\Delta c$
are objective actualities. But, their magnitudes (i.e., the spatial
and temporal lengths) are expressed merely in ratios to the defined
reference magnitudes, and each of the defined reference magnitudes
can be objectified only when it is expressed using the quantity of
ex-entity with the essential objectivity. As a result, every
physical magnitude including the spatial and temporal lengths can be
objectified only when it is expressed on the basis of the quantity
of ex-entity.

In this sense, we can conclude that the relativity related to the
spatial and temporal lengths is obtained as the result of
descriptive objectivity. In order to justify this conclusion, we
will verify, in the following Sec.III.D.4, that comparisons of the
spatial and temporal lengths corresponding to the same quantity of
ex-entity lead to the Lorentz transformation, which implies the
relativity in the spatial and temporal lengths. For all that, in
order to prevent any misunderstanding about the relativity, it is
necessary to remember that the objectivity in the descriptive
objectivity can be achieved on the basis of the quantity of
ex-entity with the essential objectivity. Given that the density of
object is dependent on the volume of object, we can see that the
afore-mentioned relative description of density results from the
relativity of spatial length.
\subsubsection{Mode of Density Change}
In view of the definition of density, the change of density can be
accomplished through two different ways -- \textit{a change of
quantity contained in a fixed volume and a change of volume occupied
by a fixed quantity}. Nevertheless, as mentioned above, since the
quantity of ex-entity has the essential objectivity, the statements
on quantity cannot be relative to each other. Hence, the change of
density induced by a movement of object cannot be accomplished by a
change of ex-entity quantity. Furthermore, the change of density
without any change of volume is incompatible with the
Michelson-Moley experiment that excluded the theory of ether from
physics. In conclusion, \textit{the change of density induced by a
movement of object is accomplished by means of not the change in the
ex-entity quantity of object but just the change in the volume of
object. In this sense, the kinetic factor represents not the change
of quantity of ex-entity but the changes of density and volume,
which are caused by the movement of object}. This conclusion enables
us to explain the relativistic contraction of length, as will be
discussed concretely in the following section.

In this sense, we can say that the acceleration of object is a
compression process with invariance of quantity in that it is
accomplished by a contraction of its volume without any change of
quantity. The kinetic energy that increases with acceleration is
therefore related not to an increase in the quantity of ex-entity
but to a compression in the volume of object. For the same reason,
the deceleration of object is an expansion process with invariance
of quantity in that it is accomplished by an expansion of compressed
volume without any change of quantity. But, a volume expanding in
the deceleration process will compress repeatedly a corresponding
quantity of space, because the space is not empty. In addition,
given the conservative characteristic of ex-entity, an expansion of
volume in the deceleration will cause a continuous propagation of
compressed density through the space. In this connection, we can
interpret the movement of matter as a wavelike propagation of
density of ex-entity and may intuitively explain the conservative
characteristics of kinetic energy and momentum. (We will discuss
quantitatively the conservative characteristics of kinetic energy
and momentum later.)

In the meantime, from the relativistic mass formula
$m(\mathbf{v})=m_0\gamma(\mathbf{v})$, an acceleration in the
special relativity is understood as a non-invariant process
accompanied with an increase of mass. In this sense, if we regard
mass as the quantity of ex-entity, the above interpretation is
incompatible with the relativistic understanding. The mass should
therefore be distinguished from the quantity of ex-entity in spite
of similarity in meaning  between them. The reason of this
discrepancy between the mass and quantity of ex-entity will be
explained on the ground of representativeness of mass in the
Sec.III.F.3.

Despite this discrepancy, considering the invariance of quantity of
ex-entity in the acceleration and deceleration, it is obvious that
the afore-mentioned interpretation of kinetic energy is compatible
with the definition of energy -- the capacity of a physical system
to do work. Owing to this compatibility, we can interpret the
relativistic formula such as $E=mc^2$, which has been experimentally
verified, in the same way as the conventional viewpoints of current
physics. In this respect, we can say, without any violation of known
empirical results, that the relativistic mass formula represents not
the increase of rest mass but the compression of rest volume.

On the other hand, given that the general relativity, it seems that
the distribution factor does not cause contradictory statements of
two observers. In this sense, \textit{the distribution factor, which
is related to the position-dependent change of density, can be
understood to represent an actual change in the quantity of
ex-entity}. Even so, the quantity of ex-entity corresponding to the
reference magnitude may be changed with the density of ex-entity,
and this density dependence of reference magnitude can be
objectified only when the reference magnitudes are described on the
basis of the same quantity of ex-entity, as discussed above. The
dependence of reference magnitudes related to the distribution
factor will be again discussed in the relation to the general
relativity, in the Sec.III.E.4.

In addition, from the preceding considerations, we can conclude
that, even in space with uniform density, the movement of object is
the only method that can change the density of object without any
change in quantity of ex-entity. That is, only the movement of
object can satisfy the relation between internal density (i.e.,
$\gamma$) and external density (i.e., $\phi$), which is required by
Eq. (17), without any change in quantity of ex-entity. This
conclusion is an explanation for the substantial reason of movement,
which was asked above.
\subsubsection{Density dependence of behavior of meter sticks and clocks}
As we have seen thus far, the relative description of quantity is
not permitted for the essential objectivity of quantity. The density
dependence of behavior of meter sticks can therefore be objectified
when it is described by a quantitative relation between lengths
corresponding to the same quantity of ex-entity. To describe this
quantitative relation, let us refer to the lengths of objects with
densities of $\rho$ and $\rho'$, which correspond to the same
quantity of ex-entity, as L and L$'$ respectively. Then, from Eq.
(3), the quantitative relation between L and L$'$ is given by
\begin{equation}
 L'=\frac{\rho}{\rho'}~L.
\end{equation}
To verify the special relativistic results, let us assume that a
difference between $\rho$ and $\rho'$ results from the relative
motion of objects. For example, if $\rho'$ denotes the density of an
object moving with velocity \textbf{v} and $\rho$ denotes that of a
stationary object, a relation between $\rho$ and $\rho'$ is written
by the above Eq. (20). For convenience, let us assume that both the
objects have the same shape and volume when both of them are at
rest, and that L and L$'$ denote the lengths of stationary and
moving objects, respectively, in the direction of motion. Here, as
is well known, the length perpendicular to the direction of motion
is independent of the motion of object\cite{Gasiorowicz2}. Hence,
from Eqs. (20) and (22), a relation between L and L$'$ is given by
\begin{equation}
 L'=\frac{L}{\gamma(\mathbf{v})}.
\end{equation}
This equation shows that the length of moving object becomes shorter
than that of stationary object in the direction of motion: as a
result, this is equivalent to the special relativistic conclusion of
length contraction.

In the meantime, we can say that the change of density related to
the kinetic factor has an anisotropic property in that the length
changes only in the direction of motion. Contrary to this,
\textit{if we consider a small region of space}, it seems that such
anisotropy of density is not found in connection with the
distribution factor. We can therefore expect that the volume of
object changes isotropically with the position of object, unlike the
case of kinetic factor. But, as far as I know, there is no
experiment for verifying a relation between distribution factor and
volume and in fact, I am not certain whether such experiment has a
physical meaning and whether it is possible.

Now, let us consider the time dilation that is another famous result
of special relativity. The temporal length can be determined by
measuring the advancing length of reference phenomenon, as discussed
in Sec.III.B.4. In addition, similar to the case of spatial length,
the density dependence of behavior of clocks can be also objectified
by making a comparison between temporal lengths corresponding to the
same quantity of ex-entity. For this comparison, at first, let us
assume that two reference phenomena occur in regions with densities
of $\rho$ and $\rho'$ and that their advancing lengths that
correspond to the same quantity of ex-entity are denoted by $L_T$
and $L_T'$, respectively. Then, from Eq. (4), a relation between
$L_T$ and $L_T'$ can be written by
\begin{equation}
 L'_{T}=\frac{\rho}{\rho'}~L_{T}.
\end{equation}
Like the case of length, let us assume that $\rho'$ denotes the
density of object moving with velocity \textbf{v} and $\rho$ denotes
the density of stationary object\footnote{This assumption is
justified by equivalence between densities of object and space,
which is written by Eq. (17) and has been mentioned in the last
paragraph of Sec.III.C.2.}. Then, since $L_T$ and $L_T'$ denote the
temporal lengths of stationary and moving clocks respectively, by
substituting Eq. (20) into Eq. (24), we have
\begin{equation}
 L'_{T}=\frac{L_{T}}{\gamma(\mathbf{v})}.
\end{equation}
Here, since $L_T$ and $L_T'$ are the reference advancing lengths,
they correspond to the numbers of ticks of clocks that are
initialized to zero; that is, $L_T$ and $L_T'$ represent frequencies
of stationary and moving clocks, respectively. Hence, Eq. (25) also
coincides with the special relativistic conclusion that a moving
clock runs more slowly than a stationary clock.

As a result, the conclusions of the present paper on behavior of
meter sticks and clocks coincide with those of special relativity.
In this sense, it is clear that the Michelson-Moley experiment,
which excluded the concept of ether from physics, can be also
explained based on the above conclusions. Nevertheless, given that
the present conclusions were obtained from the attempt of describing
the quantity of non-empty space objectively, we can say that the
Michelson-Moley experiment is compatible with the idea of non-empty
space, unlike usual interpretations. Of course, if we interpreted
the space as a kind of matter like in the ether theory, the
compatibility would be impossible as in the past 1900 or
thereabouts. But, \textit{if we interpret the space and matter as
regions of ex-entity that are classified according to density and
correlate the movement of matter with the density of space based on
objectivity and conservativeness of ex-entity, this compatibility
between the Michelson-Moley experiment and the idea of non-empty
space is possible, as discussed above}.

In conclusion, the objectivity of ex-entity (especially, in
quantity), which can be found from only the non-empty space, is the
root of all physical regularities, and moreover it is paradoxically
the ground on which the special relativity, which denied the
necessity of non-empty space, is proved valid. In particular, it is
obvious that the afore-mentioned descriptive objectivity in length
and time is one of such physical regularities and that the
invariance of space-time interval -- the keyword of relativistic
consideration -- is another way of representing the objectivity of
quantity of ex-entity.

\subsubsection{Coordinate Transformation}
Now, we will briefly discuss relations between the coordinates of an
event in primed and unprimed coordinate systems, where the primed
coordinate system is moving with a velocity v relative to the
stationary unprimed coordinate system. To avoid any unnecessary
complication, we assume for convenience that the primed and unprimed
coordinate systems have their axes parallel, that the x and $x'$
axes coincide, that the origins O and O' coincide at $t=t'=0$, and
that the direction of motion is parallel to the x and $x'$ axes as
shown in Fig. 3.
\begin{figure}
\epsfig{file=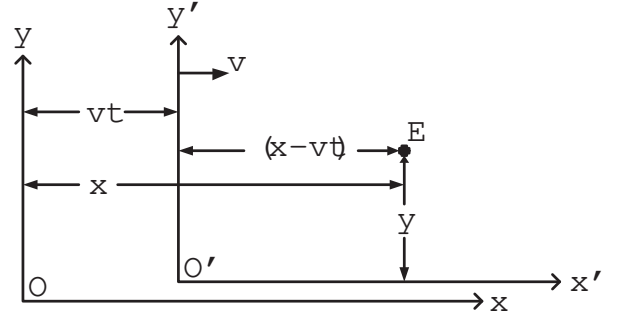, width=0.45\textwidth} \caption{Two coordinate
systems in relative motion. \label{FIG1}}
\end{figure}

Consider an event E which occurs at a point (x, y, z) at a time t in
the unprimed coordinate system. Let the primed coordinates of this
event be ($x', y', z', t'$). According to the well-known Galilean
transformation, the primed coordinates are given by
\[x'=x-\mathrm{v}t,~~y'=y,~~z'=z,~~t'=t \]
But, \textit{since the coordinates are ratios to the corresponding
reference length, they are in inverse proportion to their reference
lengths}. That is, as is well-known, the Galilean transformation is
not correct. Given this point and the relation of Eq. (23), we can
see that the primed coordinate $x'$ is equal to the multiplication
of the Galilean coordinate (i.e., x-vt) and the kinetic factor
(i.e., $\gamma$), as follows:
\begin{subequations}
\begin{equation}
 \mathit{x'}=\gamma(\mathbf{v})(\mathit{x}-\mathbf{v}t),
\end{equation}
Unlike the primed coordinate $x'$ in the direction of motion, the
primed coordinates $y'$ and $z'$, which are perpendicular to the
direction of motion, are equal to those of unprimed system because
kinetic factors in these directions are the unity. That is, we have
\begin{equation}
 \mathit{y'}=\mathit{y},~\mathit{z'}=\mathit{z}.
\end{equation}
\end{subequations}

Finally, the primed time coordinate $t'$ can be easily obtained by
conventional methods\cite{Griffiths1} \cite{Symon} that use Eq.
(26a) in order to derive the Lorentz transformation, as follows:
\begin{equation}
 t'=\gamma(\mathbf{v})\left( t - \frac{\mathrm{v}x}{c^2} \right).
\end{equation}
\subsection{Density Determination II}
\subsubsection{Equation of Motion}
From the law of motion written by Eq. (17), the ratio function M,
which is a function of position and velocity, is a time-independent
constant. Thus, we have
\begin{equation}
 \frac{d}{dt}\mathit{M}  (\mathbf{r}(t), \mathbf{v}(t))=0.
\end{equation}
Using the chain rule, this equation can be rewritten by
\begin{equation}
 \sum_{i=1,2,3}\left(\frac{\partial \mathit{M}}{\partial x_{i}} \frac{dx_{i}}{dt} +
 \frac{\partial \mathit{M}}{\partial v_{i}} \frac{dv_{i}}{dt}\right)=0.
\end{equation}
To simplify this equation, let us define gradient operators related
to position and velocity as follows:
\begin{eqnarray}
 \nabla &\equiv&
  \frac{\partial}{\partial x} \mathbf{\hat i} +
  \frac{\partial}{\partial y} \mathbf{\hat j} +
  \frac{\partial}{\partial z} \mathbf{\hat k},\nonumber\\
 ~\nabla_\mathrm{v} &\equiv&
  \frac{\partial}{\partial v_{x}} \mathbf{\hat i} +
  \frac{\partial}{\partial v_{y}} \mathbf{\hat j} +
  \frac{\partial}{\partial v_{z}} \mathbf{\hat k}.
  \nonumber
\end{eqnarray}
Using these operators, Eq. (29) is given by
\begin{equation}
 \mathbf{v}\cdot \nabla \mathit{M} + \mathbf{a}\cdot \nabla_\mathrm{v}
 \mathit{M}=0.
\end{equation}

Substituting Eq. (17) into Eq. (30), we have
\begin{equation}
\mathbf{a}\cdot \nabla_\mathrm{v} \gamma =
\frac{\gamma}{\phi}\mathbf{v}\cdot \nabla \phi.
\end{equation}
From Eq. (21), the term $\nabla_\mathrm{v} \gamma$ of Eq. (31) is
expressed by
\begin{eqnarray}
 \nabla_\mathrm{v} \gamma=\gamma^3 \frac{\mathbf{v}}{c^2}.
\end{eqnarray}
Substituting this result into Eq. (31) and then using Eq. (17) to
eliminate $\gamma$ in the resultant expression, we can obtain the
equation of motion that expresses the quantitative relation between
\textbf{a} and $\nabla\phi$, as follows:
\begin{equation}
\mathbf{v}\cdot
  \left[\mathbf{a}-\frac{c^2}{M^2} \frac{1}{\phi^3} \nabla\phi \right]=0.
\end{equation}

It is worth noting that this equation is expressed by a power per
unit mass. In addition, as mentioned in Sec.III.A, this equation
shows that a non-uniform distribution of density can lead to an
accelerative motion of object and the magnitude of acceleration is
dependent on the function of density distribution (i.e., the
distribution factor). Now, if we know the distribution factor, we
can describe the motion of object using the above Eq. (33). But the
distribution factor should be determined empirically, because it
cannot be directly measured as mentioned repeatedly above. The next
section is related to a process of determining the distribution
factor.
\subsubsection{Determination of Distribution Factor}
At first, let us consider a gravitational field that is one of the
most important fields in the history of physics. As is known, the
gravitational field is written by
\begin{equation}
\mathbf{g}=-\frac{Gm}{r^2}\hat{\mathbf{r}},
\end{equation}
where G denotes the gravitational constant and m does the mass of
source object that generates the gravitational field. Meanwhile,
given a vector equation
$\mathbf{A}\cdot\mathbf{B}=\mathbf{A}\cdot\mathbf{C}$, two vectors
$\mathbf{B}$ and $\mathbf{C}$ should generally satisfy other vector
equation $\mathbf{A}\times\mathbf{B}=\mathbf{A}\times\mathbf{C}$ so
that $\mathbf{B}$ is equal to $\mathbf{C}$. We cannot therefore
remove the term $'\mathbf{v}\cdot'$ from Eq. (33) freely. But, in
the above Eq. (33), if $\mathbf{a}$ is parallel with $\nabla\phi$,
the equation
\begin{equation}
\mathbf{a}=\frac{c^2}{M^2} \frac{1}{\phi^3} \nabla\phi.
\end{equation}
holds as the solution of Eq. (33). To satisfy this requirement, let
us assume that $\nabla\phi$ is radial (i.e., $\phi$ is isotropic).
It is empirically obvious that this assumption is approximately
valid for many physical situations. Thus, substituting Eq. (34) into
Eq. (35), we have
\begin{equation}
\frac{c^2}{M^2} \frac{1}{\phi^3} \frac{d\phi}{dr}=-\frac{Gm}{r^2}.
\end{equation}
Solving this differential equation with respect to $\phi$ and r, we
have
\begin{equation}
 \phi(\mathbf{r}) =\left[\frac{1}{\phi^2_0}- \frac{2GmM^2}{c^2}
        \left(\frac{1}{r}-\frac{1}{r_0} \right)\right]^{-1/2}.
\end{equation}
For convenience, let us select the infinity for the reference
position $r_0$ where the reference density is defined. Then, since
the reference density is unity as mentioned above, Eq. (37) can be
written in the simple form as follows:
\begin{equation}
 \phi(\mathbf{r}) =\left[1- \frac{2GmM^2}{rc^2}\right]^{-1/2}.
\end{equation}
Given that the value of M is dependent on how to select the
reference position as mentioned above, we can properly select the
reference position such that M becomes the unity. For M=1 like this,
the distribution factor of Eq. (38) seems to be related to the
Schwarzschild solution of the general relativity. It is worth noting
that both the Eq. (38) and the Schwarzschild solution are related to
a static field having a spherical symmetry. In this respect, if the
kinetic factor is a keyword of special relativity, we can say that
the distribution factor of Eq. (38) for M=1 is a keyword for general
relativity. In the following sections, we will examine the meaning
of the constant M and then discuss behaviors of meter sticks and
clocks under the gravitational field.
\subsubsection{Meaning of the Constant of Motion}
Substituting Eqs. (21) and (38) into Eq. (17) for M, we have
\begin{equation}
 M(\mathbf{r}, \mathbf{v}) =\left[1- \frac{\mathrm{v}^2}{c^2}
                +\frac{2Gm}{rc^2}\right]^{-1/2}.
\end{equation}
As in the relativistic analysis of the famous formula $E=mc^2$,
expanding the right side of Eq. (39) and then multiplying the
resultant expansion by $m_0c^2$, we have
\begin{eqnarray}
 Mm_0 c^2 =m_0 c^2 +\frac{m_0\mathrm{v}^2}{2}-\frac{Gm_0m}{r}
  + O(\mathrm{v}^4, \frac{1}{r^2}).
\end{eqnarray}
The first term of right side of this equation corresponds to the
energy of rest mass, and the second and third terms do the kinetic
and potential energies of the classical mechanics, respectively. The
remaining terms of right side can be interpreted as the relativistic
kinetic and potential energies. In conclusion, we can say that the
constant of motion M represents (total mechanical energy)/$m_0c^2$,
which is expressed as a function of position and velocity.

\subsubsection{Distribution Factor dependence of behavior of meter
sticks and clocks}

As have been discussed in Sec.III.D.3, the density dependence of
behavior of meter sticks and clocks can be objectified by the
comparisons of spatial and temporal lengths corresponding to the
same quantity of ex-entity. In that section, we have also obtained
Eqs. (22) and (24) for such objective comparisons of spatial lengths
and of temporal lengths, and the density dependence of behavior of
meter sticks and clocks has been discussed in connection with the
kinetic factor. But, unlike the kinetic factor, the change of
distribution factor represents not a change of volume occupied by
the same quantity of ex-entity but an actual variation of quantity
of ex-entity contained in the unit volume, as mentioned in
Sec.III.D.2. In this sense, it is expected that a change of density,
which is caused from a positional change, does not result in an
anisotropic change of length (e.g., the length contraction in the
moving direction induced by a movement); that is, a change of length
related to the distribution factor seems to be isotropic. Given this
isotropic property, if the distribution factor of space is expressed
by Eq. (38), a relation between two lengths L and L$'$ can be
expressed by
\begin{equation}
 L' =L \left(1-\frac{2Gm}{rc^2}\right)^{1/6},
\end{equation}
where L and L$'$ denote spatial lengths that are defined at the
reference position $\mathbf{r}_0$ and an arbitrary position
\textbf{r}, respectively, and correspond to the same quantity of
ex-entity. But if ever there was unknown anisotropy, a relation
between lengths in the anisotropic direction may be written by
\begin{equation}
 L' =L \sqrt{1-\frac{2Gm}{rc^2}}.
\end{equation}
At all event, given that a change of distribution factor represents
an actual variation in the quantity of ex-entity, we can say that L
and L$'$ in Eq. (41) or Eq. (42) represent not real lengths but just
the magnitudes of length corresponding the same quantity of
ex-entity (hereinafter, length-values). On the contrary, a variation
of length-value related to the kinetic factor is necessarily
accompanied by an actual change of length due to the invariance of
quantity of ex-entity, as we have seen above.

In the meantime, we can identically apply the argument made in
Sec.III.D.3 to a distribution factor dependence of temporal length.
That is, if the distribution factor of space is given by Eq. (38)
and two clocks are disposed at the reference position $\mathbf{r}_0$
and an arbitrary position $\mathbf{r}$, respectively, a relation
between temporal lengths, which are expressed by the two clocks, can
be obtained from Eqs. (24) and (38) as follows:
\begin{equation}
 L'_{T} =L_{T}~ \sqrt{1-\frac{2Gm}{rc^2}},
\end{equation}
where $L_T$ denotes a temporal length of one clock at $\mathbf{r}_0$
and $L_T'$ denotes that of the other clock at \textbf{r}. Here, as
mentioned above, a frequency of light corresponds to a temporal
length of clock. Thus, the distribution factor dependence of
temporal length expressed by Eq. (43) can be quantitatively verified
by measuring how a frequency of light, which is generated at
$\mathbf{r}_0$, changes at \textbf{r} or vice versa, as suggested in
the theory of general relativity\cite{Weinberg}. As a result, the
above Eq. (43) coincides with the gravitational time dilation
predicted by the theory of general relativity, and therefore, we can
say that our argument is compatible with the theory of general
relativity at least within the scope that has been discussed so
far\footnote{The general theory of relativity have quantitatively
predicted several amazing phenomena, such as precession of mercury,
deflection of light, and radar echo delay, and these predictions
have been verified experimentally so far. But, since these
predictions are based on tensor analysis, which is difficult for me,
I have not dealt with these phenomena quantitatively. I hope that
these general relativistic subjects will be further studied by
readers.}.
\subsection{Further Consideration}
\subsubsection{Distribution Factor and Stability of Matter}
As discussed in Sec.II.C, the matter is merely a local part whose
existence can be perceived amid the universal distribution of
object. Nevertheless, the mechanical description based on the matter
has been successful as shown in the classical mechanics. Owing to
this success of mechanical description, we may say that the matter
is a representative that symbolizes the universal distribution of
object. But, given that other parts of object whose existence cannot
be perceived may still exist outside the matter, it is necessary to
explain reasons that the mechanical description can be successful
and that the matter can be regarded as the representative of
universally distributed object.

Furthermore, the law of motion written by Eq. (17) governs motions
of all points of universally distributed object, because the law of
motion is the universal statement that can be applied for every
case. We should therefore describe motions of not only the local
part, which can be observed as the matter, but also each and every
part of object, in order to give a complete account of phenomena. In
this sense, the mechanical description based on the matter is not
complete, though useful and effective.

Nevertheless, since we should consider motions of infinite points
for this universal description, we are inevitably confronted with
complexity and difficulty in the mathematical description of
physical world. In addition, the universal distributions of objects
result in a superposition or an entanglement of objects over the
whole universe, which causes a difficulty in distinguishing a test
object from a source object. In some respect, this difficulty of
distinction casts doubt on whether the matter is proper as the
representative of universally distributed object.

But, the matters such as electrons and protons are remarkably
physically stable, as is well known. I believe that this stability
of matter is a clue to the solution of afore-mentioned problems.
That is, \textit{owing to the strong stability of matter, we can
conjecture that the distribution factor of object has a special
mathematical structure that ensures the stability of matter, and
that such mathematical structure of distribution factor is preserved
anyhow}. In this case, the matter can be justified as the
representative of universally distributed object and the complexity
and difficulty in mathematical description can be alleviated by the
mechanical description based on the matter. It is also possible to
distinguish a test object from a source object, because the density
distributions of test and source objects will be preserved
independently. In analogy, even though two water waves generated at
different positions are superposed at many positions, they can be
independently described by different wave equations, because they
propagate independently.

In the meantime, given that the distribution factor of source object
written by Eq. (38) is not homogeneous, we can see that respective
parts of test object will be differently accelerated with each other
depending on their positions. This position-dependent acceleration
of test object inevitably leads to deformation of distribution
factor of test object. But, from the above conjecture, we can expect
that a distribution factor of object will be restored to preserve
the stability of object. I believe that this restoring process can
be correlated with electromagnetic or quantum mechanic phenomena,
such as the radiation caused by the transition of electron. Of
course, to justify this belief, we should further study for
answering to remaining issues, such as the reason that the
distribution factor of matter has the form of Eq. (38) and the
mathematical particularity of distribution factor. It seems that the
remaining issues are closely related to the quantum mechanics.

\subsubsection{Force and Field}
The concept of force in Newtonian mechanics is definitely useful for
the first step of analyzing unknown phenomena, but it is just a
concept based on the representativeness of matter. Though the
representativeness of matter can be successfully used for
alleviating the difficulties in physical description, the success of
mechanical description based on the representativeness of matter
cannot justify the groundless belief that all physical substances of
object are contained in the local region referred to as the matter.
In this sense, we cannot say that the force, which describes only a
motion of matter, is always useful for an investigation of the
physical world. In fact, if we can fully know density functions of
objects and calculate acceleration at every point of test object
from the density functions, there is no need to depend on the
representativeness of matter anymore. That is, further complete
information on the physical world can be obtained from not the
concept of Newtonian force but the acceleration at every point of
object.

Contrary to the force, the physical field such as the gravitational
field and the electric field provides us with physical information,
which is related to the acceleration, at every point of the whole
universe, as is well known. That is, the acceleration at every point
may be obtained from the physical field. But, the physical
properties of field, such as magnitude and direction, are determined
only by the source object (especially, its density distribution),
while the force or the acceleration is an interaction between the
test object and the field. In this sense, we cannot identify the
field with the real acceleration at every point of test object. To
calculate the real acceleration of test object, we should know not
only the density distribution of source object, from which the field
is generated, but also a physical property of test object. Of
course, the property of test object that affects the acceleration
should be naturally related to the density of test object. In
conclusion, \textit{we should know the acceleration at every point
of test object in order to have a better understanding of the
physical world, and we should consider both the densities of test
and source objects to calculate this acceleration.}

In the meantime, given the Galileo's famous experiment in Pisa, it
appears that the magnitude of acceleration of test object is
independent of the absolute values of density of test object.
Contrary to this, as we shall see in following section, the
direction of acceleration is dependent on the density directions of
test and source objects. Here, the density direction is defined as a
parameter for indicating whether the density of object is larger or
lesser than unity, and can be understood as the sign of electric
charge as will be argued later.

\subsubsection{Charge, Ex-entity and Mass}
In the above Sec.III.E.2, we have seen that the gravitational field
can be generated from the density distribution of ex-entity written
in Eq. (38). As is well known, there are electromagnetic, weak, and
strong interactions besides the gravitational interaction in nature.
Considering the singleness of ex-entity, we cannot however introduce
extra distribution factors for each of the remaining fields. That
is, \textit{if we can know the density function of source object
completely and correctly, the remaining fields as well as the
gravitational field should be obtained from the density function of
source object\footnote{I will discuss the electromagnetic field
later in this paper, but not the strong and weak fields beyond my
ability. In addition, not only the density function of source object
but also that of test object may be needed to describe some
interaction.}.}

At first, let us glance over the electric force. As is well known,
the electric force is analogous to the gravitational force in that
the intensity of force varies inversely as the square of distance,
but they are different from each other in that the direction of
electric force is dependent on the sign of electric charge. In this
sense, it is necessary to examine a relationship between the density
of ex-entity and the sign of electric charge before having a
concrete discussion on the electric field itself.

For this examination, to begin with, let us consider how to express
the afore-mentioned density direction of object. Once, let us
re-write the distribution factor of source object, written in Eq.
(38), in a general form as follows:
\begin{equation}
 \phi_{s}(\mathbf{r})=  \left[\sqrt{1+\frac{2q_s k_s}{rc^2}}~\right]^{-1}.
\end{equation}
where $k_{s}$ is a parameter that characterizes the distribution
factor of source object. Considering the above Eq. (35), if $q_{s}$
is the sign of electric charge, the characteristic parameter $k_s$
will be 1/$4\pi\epsilon_{0}$ for the electric force: $k_{s}$ is a
positive constant. Hence, the value of $\phi_s$ is 1 or less at
every point for positive $q_s$ and is 1 or more in the most region
(i.e., $2k/c^2\leq r\leq\infty$) for negative $q_s$. As a result,
the parameter $q_s$ included in Eq. (44) represents the density
direction of source object, which was defined in the previous
section.

Next, let us verify that the parameter q can be understood as the
sign of electric charge. For this, substituting Eq. (44) into Eq.
(35), we can express a field \textbf{E} generated from the
distribution factor of Eq. (44) as follows:
\begin{equation}
 \mathbf{E}=q_s\frac{k_s}{r^2} \hat{\mathbf{r}}.
\end{equation}
At this time, since the Eq. (45) is obtained only from the
distribution factor of source object, it corresponds to not a
measurable acceleration field but just some physical field (in fact,
the electrostatic field). Of course, this is identical to the case
of gravitational interaction. But, for the gravitational
interaction, an acceleration of test object is independent of its
density or its density direction as known empirically. The
gravitational field can therefore be expressed as the acceleration
field of test object, for convenience. Contrary to this, for the
electric interaction under consideration, the acceleration acting on
the test object is dependent not only on the density of source
object but also on that of test object. (Strictly speaking, it
depends on the density directions of objects-the signs of electric
charges.) In this respect, we need to consider the density
distribution of test object besides that of source object written in
Eq. (44). For this, let us write the distribution factor of test
object in the same form as that of source object, as follows:
\begin{equation}
 \phi_t(\mathbf{r})=
 \left[\sqrt{1+\frac{2q_t k_t}{(r-r_t)c^2}}~\right]^{-1}.\nonumber
\end{equation}
where $q_t$, $k_t$ and $r_t$ denote the density direction, the
characteristic parameter and the position of distribution center
respectively of test object. In addition, let us assume\footnote{In
fact, this assumption is introduced because of its likelihood; I
cannot concretely explain its ground. Further discussion is needed.}
that the acceleration of test object in the field of Eq. (45) can be
written by the product of $q_t$ and \textbf{E}, as follows:
\begin{equation}
 \mathbf{a}=q_t q_s\frac{k_s}{r^2} \hat{\mathbf{r}}.
\end{equation}
In this case, the q-values of source and test objects determine the
direction of acceleration, because $k_s$ is positive. That is, the
acceleration of test object is repulsive in the case of the same
q-values and is attractive in the case of different q-values. This
feature of direction of acceleration coincides with that of the
known electric force. We can therefore conjecture that the q-value
denotes the sign of electric charge.

From this result, we can say that for the electric force, the q is a
sign of electric charge and the characteristic parameter k is
1/$4\pi\epsilon_{0}$, while for the gravitational force, the
characteristic parameter k is -Gm/q regardless of q. But, given that
extra distribution factors cannot be introduced for several
interactions as stated above, we can conclude that at least one of
the gravitational and electric interactions is a secondary effect. I
do not know what the primary one is. But given that the
gravitational interaction dominates only the electrically neutral
world and is manifestly weaker than the electric interaction, it
seems that the gravitational interaction is the secondary one. Of
course, all attempts at explaining the gravitational interaction
based on the electric interaction have failed up to
now\cite{Feynman1}. It is however likely that there is no attempt
based on the idea presented in this paper. Hence, further studies
are needed to verify this assumption.

As for another issue, let us discuss a relation between the quantity
of ex-entity and the mass. From the definition of density, we should
consider the volume of object in order to calculate a total quantity
of ex-entity of object. That is, given the universal distribution of
object, the total quantity of object can be obtained by integrating
the distribution factor over the whole space. But, since the
distribution factor $\phi$ is close to unity in the most regions
regardless of q-value, such integral over the whole space does not
converge. This approximate unity in the value of distribution factor
may however be interpreted in connection with the fact that all
existing objects are superposed and distributed over the whole
space. In this sense, let us introduce a proper distribution factor
of object that is defined as a function of subtracting unity, which
can be apprehended as a consequence of superposition of
distributions of other objects, from the distribution factor of
object, as follows:
\begin{equation}
 \phi_0(\mathbf{r})\equiv  \phi(\mathbf{r})-1 =
 \left[\sqrt{1\pm \frac{2k}{rc^2}}~\right]^{-1} -1.
\end{equation}

Someone may say that Eq. (47), which is approximately zero in the
most regions, is relevant to the concept of mass, because the mass
is a concept based on the belief that the space is a vacuum. But,
even if the unity problem is taken into consideration, an integral
of Eq. (47) over the whole universe does not converge as before.

In fact, the total ex-entity quantity of object cannot be equal to
the mass of object, because the mass is just one of representative
magnitudes characterizing the object. Concretely, the mass is
defined as a ratio to acceleration in the Newtonian definition of
force, and the force is one of concepts based on the
representativeness of matter as mentioned above. In addition, the
mass is based on the groundless beliefs that it is contained in the
local region referred to as the matter and the space is a vacuum. In
this sense, we can say that the mass is just a representative
magnitude, as mentioned above. And, the mean value is no more equal
to the total than the mass, which is a representative magnitude, is
equal to the total ex-entity quantity of object. This situation is
identical to the amount of electric charge that is expressed by the
unit of Coulomb (C). As a result, even if we can calculate the total
ex-entity quantity of object, the result may be not equal to the
mass or the charge amount in general. Considering this point, the
afore-mentioned q-value - density direction - is not the charge
amount of object but just the density direction of object. This
should be remembered to avoid unnecessary confusion in the following
sec. III.F.6.

\subsubsection{Singularity}
For q=-1, Eq. (44) is singular at $r=2k/c^2$ that corresponds to the
Schwarzschild radius. This singularity may be analyzed in connection
with the definition of density -- the quantity of ex-entity
contained in the unit volume. To put it concretely, the singularity
of density seems to be related to the fact that information on
volume vanishes in a zero-dimensional point.

From the above discussion on the kinetic factor, the density of
object becomes infinity as the speed of object reaches that of
light. But, as mentioned above, the increase of kinetic factor is
the process of compression with the invariance of quantity. Hence,
even if the density of ex-entity is infinite at an arbitrary point,
the quantity of ex-entity cannot be infinite at the point. In
particular, given that the infinite quantity may literally fill the
whole universe with infinite quantity, it is obvious that the
quantity of point is physically impossible to be infinite. The
infinity of density related to the distribution factor is equivalent
to this, because the distribution factor should be determined from
the kinetic factor as mentioned in sec. III.E.

This difference between density and quantity is caused by the fact
that, from the definition of density, the density is calculated
based on the unit volume. That is, whenever a finite quantity
contained in a non-zero volume is compressed into a zero-dimensional
point, the resultant density becomes infinity regardless of initial
volume. In this sense, \textit{if we wish to calculate an ex-entity
quantity of point having a specific density objectively, we should
consider the unit volume, which is used as a reference for
calculating the density}.

In the meantime, I thought that the Archimedes' idea could be used
for a calculation of ex-entity quantity, but I found recently that
my calculation was based on a fatal mistake. The subject on
singularity will therefore be not discussed any more in this paper.
Nevertheless, it is worth noting that the singularity of Eq. (47) is
distinguished from singularities in conventional field physics. That
is, the singularity in the field physics is generally related to the
potential, but the potential is just an abstract concept that is
generated from the formalistic approach based on the mathematics.
Hence, this formalistic approach hinders us from understanding
physical meanings of potential or singularity. Contrary to this, the
singularity of Eq. (47) is related to the function of ex-entity
density, and as mentioned above, the ex-entity density is the
concept that can be understood intuitively and has
volume-dependence. In this sense, if we take the unit volume into
consideration, the singularity of density may be possibly understood
in a substantial level.

\subsubsection{Spin and Size of Particle}
Up to now, we have discussed the situation in which the density
distribution of source object is static. But, given that an object
is distributed over the whole space, it is clear that the density
distribution of object can be changed in various ways. In this
section, we will discuss the rotation of object - the revolution of
object on its own center-, and in the next section, we will consider
the influences of the movement of test object caused by the movement
of source object. But, in this paper, we do not consider the orbital
motion of object in which the object revolves around some point or
axis that is deviated from its own center.

As is well known, Stern-Gerlach's experiment has shown the fact that
the electron has a spin angular momentum, but the known spin angular
momentum of electron is incompatible with the maximum size of
electron that is calculated from scattering experiments. In
addition, if we regard the electron as a point particle, due to the
degree of freedom, we cannot explain the reason that the electron
has a finite spin angular momentum\cite{Beiser}. Owing to these
contradictions, the modern physics explains that the electron is not
a classical particle but a quantum mechanical
particle\cite{Eisberg}. Frankly speaking, I am ignorant of Dirac's
relativistic quantum mechanics that is known as providing the
substantial explanation for the spin of electron. Owing to my
ignorance, I suspect that the modern physical explanation based on
the quantum mechanics means not a solution of above contradictions
but only a success of mathematical description of phenomena.

At all event, if the electron is distributed universally as
mentioned above, the contradiction of classical models related to
the spin of electron can be solved at least qualitatively. Firstly,
it is obvious that the contradiction related to the degree of
freedom disappears, because the universally distributed electron has
the degree of freedom four and over. Of course, given the universal
distribution of electron and the relativistic limit of speed, it is
obvious that the electron is not a rigid body whose all parts
revolve with the same angular velocity. Accordingly, the degree of
freedom of electron becomes infinity actually. Nonetheless, we can
conjecture that the electron has a finite degree of freedom in the
macroscopic aspect, because the afore-mentioned constraint on the
density distribution, which is required for the stability of matter,
serves to reduce the degree of freedom of electron. I believe that
the constraint is connected with the quantum numbers of particle
including the spin angular momentum. As a result, a remaining
contradiction is the inconsistency between the spin angular momentum
of electron and the maximum size of electron.

To solve this inconsistency, it is necessary to understand that the
size of particle is a concept based on the classical viewpoint of
matter that classifies matter and space according to the substantial
difference in kinds. According to the classical viewpoint of matter,
there is a substantial boundary surface between the matter and the
space to separate being and nothing, as mentioned above, and all
physical contents related to the matter are contained in the
internal region of the boundary surface. Here, the size of particle
is a concept based on the belief that there is the substantial
boundary surface, and it means generally a radius of the internal
region of boundary surface. But, it seems that the classical
viewpoint of matter is only originated from usual experience,
because there is no proof that can justify the classical viewpoint
of matter. That is, there is no evidence that the substantial
boundary surface exists between being and nothing. In this sense, we
need not to have a deep attachment to the concept of particle size.
(I believe that the duality of matter and wave, which is the origin
of quantum mechanics, is closely related to nonexistence of
substantial boundary surface, though it will be not discussed here.)

If we exclude the belief in the substantial boundary surface, we can
also solve the inconsistency between the spin angular momentum of
electron and the maximum size of electron. That is, it can be
understood that the measured spin angular momentum of electron
results from not a rotation of local region but a revolution of
every part of universally distributed electron. In this case, we can
avoid the contradictory conclusion that the localized electron
should be rotated with the angular speed, which exceeds the velocity
of light, to satisfy the measured spin angular momentum of electron.
Meanwhile, given that the size of electron is determined from the
calculation based on scattering experiments, it can be concluded
that the size of electron corresponds to not a size of imaginary
ball that contains all physical contents of electron but just an
impact parameter that is dependent on the kinetic energy of
electron. This is because the scattering experiment does not prove
that all physical contents related to the electron are confined in
the internal region of measured impact parameter. The idea that the
localized electron ball exists is just a groundless belief based on
the classical viewpoint of matter, as explained above.

In the meantime, if someone particularly wishes to define the size
of particle, the size of region having a peculiar density seems to
be a good criterion for defining the size of particle. For example,
we can define the size of electron as that of region with a zero
density; from Eq. (44), we can say that the electron is a point
particle of which radius is zero. Similarly, the size of proton can
be defined as that of region with an infinite density; in this case,
the maximum radius of proton can be given by an electrical
Schwarzschild radius defined as follows:
\begin{equation}
 r_{proton}=\frac{2e}{4 \pi \epsilon_0 c^2} \frac{e}{m_p}\simeq~3.1\times10^{-18}m.
\end{equation}
For all that, it is obvious from above arguments that these
localized regions do not contain all contents of electron or proton.
\subsubsection{Electromagnetism}

It is obvious that a movement of test object caused by a non-static
source should be different from that caused by a static source.  In
this section, we will discuss quantitatively a movement of test
object under a moving source with a non-uniform density
distribution. For such discussion, we need to express respective
densities of objects by using their distribution and kinetic factors
and to obtain an equation connecting quantitatively them, similar to
the argument of Sec. III.C.1.

At first, let us suppose that, at an arbitrary position \textbf{r},
test and source objects move with velocities of \textbf{v} and
\textbf{u} respectively relative to a fixed observer. Then, this
observer will express the densities of test and source objects,
$\rho_1$ and $\rho_2$, in terms of the products of the reference
density, respective distribution factors and respective kinetic
factors, as follows:
\begin{subequations}
\begin{eqnarray}
  \rho_{1}=\rho_0 \phi_1(\mathbf{r})\gamma_1 (\mathbf{v}(\mathbf{r})),
  \\
  \rho_{2}=\rho_0 \phi_2(\mathbf{r})\gamma_2 (\mathbf{u}(\mathbf{r})).
\end{eqnarray}
\end{subequations}
Here, it should be remembered that test and source objects are
distributed over the whole universe as stated above, and that
\textbf{v} and \textbf{u} denote respective velocities of
one-points, which are positioned at \textbf{r}, of test and source
objects. Meanwhile, considering this continuum-like property of
objects and the aforementioned stability of matter, we can assume
that, in most classical phenomena, a velocity of one-point is nearly
equal to those of neighboring points. Hence, we will assume that
\begin{subequations}
\begin{eqnarray}
  \nabla \gamma_1(\mathbf{v})\sim 0,~
  \nabla\mathbf{v}\sim 0,~
  \nabla\cdot\mathbf{v}\sim 0,~
  \nabla\times\mathbf{v}\sim 0,
  \\
  \nabla \gamma_2(\mathbf{u})\sim 0,~
  \nabla\mathbf{u}\sim 0,~
  \nabla\cdot\mathbf{u}\sim 0,~
  \nabla\times\mathbf{u}\sim 0.
\end{eqnarray}
\end{subequations}
(Though it is not discussed in this article, if we want to expand
our argument toward the region of quantum mechanics, it seems that
the above neglecting needs to be reconsidered.)

In this case, by generalizing the argument of Sec. III.C.1, we can
obtain a generalized equation of motion from Eqs. (49a) and (49b),
as follows:
\begin{eqnarray}
 M(\mathbf{r}(t), \mathbf{v}(t), \mathbf{u}(t))&\equiv &
 \frac{\gamma_1 (\mathbf{v}(t))}{\phi_2 (\mathbf{r}(t))\gamma_2 (\mathbf{u}(t))}\nonumber\\
 &=&\frac{1}{\phi_2 (\mathbf{r}(0))\gamma_2(\mathbf{u}(0))}~(const),
\end{eqnarray}
where M denotes the constant of motion depending of \textbf{r},
\textbf{v} and \textbf{u}. But, contrary to the case of Sec.
III.E.1, a time derivative of Eq. (51) includes an additional terms,
which hinders from writing an acceleration of test object to a form
of explicit function. Owing to such difficulty, I will adopt other
approaching method in the following argument\footnote{The properness
of the following assumptions should be verified from further
research for overcoming this difficulty.}.

Specifically, I will assume that a total acceleration of test object
$\mathbf{a}_{t}$ is equal to the sum of a distribution acceleration
$\mathbf{a}_{d}$, which results from the \emph{non-uniformity} of
density distribution of source, and a kinetic acceleration
$\mathbf{a}_{k}$, which results from the \emph{movement} of source.
That is,
\begin{equation}
\mathbf{a}_{t}=\mathbf{a}_{d}+\mathbf{a}_{k}.
\end{equation}
In addition, I will assume that an acceleration effect related to
the movement of source can be completely expressed by the kinetic
acceleration $\mathbf{a}_{k}$: that is, the distribution
acceleration $\mathbf{a}_{d}$ is independent of the movement of
source object. (Strictly speaking, there is no sufficient ground for
these assumptions, and therefore, my research related to the
electromagnetism is incomplete yet. For all that, considering the
following successful results obtained from these assumptions, I
think that they disclose one aspect of a perfect theory that may be
obtained from a thoughtful consideration: that is, it seems that
they can be regarded as at least one of considerable theoretical
models.) Meanwhile, as a matter of convenience, we leave the
retarded time out of consideration in the following discussion.

Now, let us calculate the distribution acceleration
$\mathbf{a}_{d}$. If the distribution acceleration $\mathbf{a}_{d}$
is independent of the movement of source as assumed above, it may be
similar in form to Eq. (35), which expresses the acceleration of
test object under the case of static source, as follows:
\begin{equation}
\frac{c^2}{M^2 (\phi_2\gamma_2)^3} \nabla(\phi_2\gamma_2),\nonumber
\end{equation}
where M is the constant of motion given by Eq. (51). However, as
explained in Sec. III.F.3, this physical quantity corresponds to not
a real acceleration of test object, which can be measured directly,
but only a field generated by the source object. As explained in
Sec. III.6.3) and 4), for the electromagnetic phenomena, the real
distribution acceleration of one point, positioned at \textbf{r}, of
test object can be given by the product of the density direction
(i.e., the sign of electric charge) of test object $q_1$ and the
field generated by the source object, which is given by the above
equation, as follows:
\begin{equation}
\mathbf{a}_{d}=q_1 \frac{c^2}{M^2 (\phi_2\gamma_2)^3}
\nabla(\phi_2\gamma_2).
\end{equation}

Now, let us calculate the kinetic acceleration $\mathbf{a}_{k}$. For
this, let us define the momentum factor as the product of the
distribution factor, the kinetic factor and the velocity at each
points of object. According to this definition, the momentum factors
of test and source objects $\mathbf{p}_1$ and $\mathbf{p}_2$ can be
written by
\begin{subequations}
\begin{eqnarray}
 \mathbf{p_1}= \phi_1(\mathbf{r})\gamma_1(\mathbf{v})\mathbf{v}
             \equiv \psi_1(\mathbf{r}, \mathbf{v})\mathbf{v},\\
 \mathbf{p_2}= \phi_2(\mathbf{r})\gamma_2(\mathbf{u})\mathbf{u}
             \equiv \psi_2(\mathbf{r}, \mathbf{u})\mathbf{u}.
\end{eqnarray}
\end{subequations}
where $\psi_1$ and $\psi_2$ denote the ratios of densities of test
and source objects, respectively, to the reference density $\rho_0$
and can be defined as the product of distribution and kinetic
factors at \textbf{r} as expressed in the above equations.

In the meantime, to know an effect on the movement of test object
caused by the moving source, we need to obtain the equation
connecting quantitatively the two momentum factors, as mentioned
above. However, such equation is unknown for the present, because
the momentum factor was not derived but introduced. That is, it is
unclear whether the well-known conservation law of momentum can be
applied for the situation under discussion. Therefore, for a
generalized discussion, let us assume that the time derivative of
the algebraic sum of momentum factors is equal to an unknown vector
\textbf{X}. Here, for the electromagnetic phenomena, we should take
the density directions of objects into consideration, as explained
above. Therefore, this assumption may be expressed by an equation
with a parameter Q that depends on the density directions of source
and test objects, as follows:
\begin{equation}
 \frac{d}{dt}\left(\mathbf{p}_1+Q\mathbf{p}_2\right) = \mathbf{X}.
\end{equation}
We will discuss a relation between the parameter Q and the density
direction $q_1$ below.

Let us calculate $d\mathbf{p}_1/dt$ and $d\mathbf{p}_2/dt$,
respectively. Given the necessity of density-based description
discussed in section III.A.1, if we want to express a movement of
test object, it is necessary to describe first a movement of one
point of test object with a specific constant density. For this, let
us select the specific point of test object whose the magnitude of
distribution factor is $\phi_{10}$ and calculate Eq. (55) along a
trajectory of this point. Then, from Eq. (54a), the time derivative
of momentum factor of selected point is given by
\begin{equation}
 \frac{d\mathbf{p}_1}{dt}=\phi_{10}\left[\frac{d\gamma_1(\mathbf{v})}{dt}\mathbf{v}+\mathbf{a}\gamma_1(\mathbf{v})\right],
\end{equation}
where \textbf{a} denotes the time derivative of velocity of selected
point of test object. Since the kinetic factor is the function of
velocity, which is given by Eq. (21), the $d\gamma_1/dt$ in Eq. (56)
can be given by
\begin{equation}
   \frac{d\gamma_1(\mathbf{v})}{dt}=\mathbf{a}\cdot\nabla_\mathrm{v}\gamma_1(\mathbf{v}).
\end{equation}
where $\nabla_\mathrm{v} \gamma_1 (\mathbf v)$ denotes the velocity
gradient of kinetic factor of selected point of test particle, which
is equal to Eq. (32). Therefore, Eq. (56) can be written by
\begin{equation}
   \frac{d\mathbf{p}_1}{dt}=
   \phi_{10}\gamma_1\left[\frac{\gamma^2_1}{c^2}(\mathbf{a}\cdot\mathbf{v})\mathbf{v}+\mathbf{a}\right].
\end{equation}
Since the $(\mathbf{a}\cdot\mathbf{v})\mathbf{v}$ equals to
$\mathbf{v}\times(\mathbf{v}\times\mathbf{a})+\mathrm{v}^2\mathbf{a}$
by the BAC-CAB rule, Eq. (58) can be written by
\begin{equation}
   \frac{d\mathbf{p}_1}{dt}=
   \phi_{10}\gamma_1\left[\mathbf{a}\left(1+\frac{\mathrm{v}^2}{c^2}\gamma^2_1\right)+
   \frac{\gamma^2_1}{c^2}\mathbf{v}\times(\mathbf{v}\times\mathbf{a})\right].
\end{equation}

And since $(1+\mathrm{v}^2 \gamma^2_1/c^2)$ equals to $\gamma^2_1$
by Eq. (21), Eq. (59) can therefore be written by
\begin{equation}
   \frac{d\mathbf{p}_1}{dt}=
   \phi_{10}\gamma^3_1\left[\mathbf{a}+\frac{1}{c^2}~\mathbf{v}\times(\mathbf{v}\times\mathbf{a})\right].
\end{equation}

Now, let us calculate the term of $d\mathbf{p}_2/dt$ in Eq. (55).
Given that an action-at-a-distance is impossible, to describe the
movement of selected point of test object, we should calculate the
time derivative of momentum factor of source object along the
trajectory of selected point of test object; that is, the term of
$d\mathbf{p}_2/dt$ should be calculated as the convective
derivative, as follows:

\begin{equation}
   \frac{d\mathbf{p}_2}{dt}=\frac{\partial\mathbf{p}_2}{\partial t}
   +(\mathbf{v}\cdot\nabla)~\mathbf{p}_2,
\end{equation}
where \textbf{v} in Eq. (61) denotes the velocity of selected point
of test object. Here, using the known vector identity
$(\mathbf{v}\cdot\nabla)\mathbf{p}_2=\nabla(\mathbf{v}\cdot\mathbf{p}_2)-\mathbf{v}\times(\nabla\times\mathbf{p}_2)
-{\mathbf{p}_2}\times(\nabla\times\mathbf{v})-(\mathbf{p}_2\cdot\nabla)\mathbf{v}$,
Eq. (61) can be written by
\begin{eqnarray}
   \frac{d\mathbf{p}_2}{dt}=\frac{\partial\mathbf{p}_2}{\partial t}
   +\nabla(\mathbf{v}\cdot\mathbf{p}_2)-\mathbf{v}\times(\nabla\times\mathbf{p}_2)
   \nonumber\\
   -{\mathbf{p}_2}\times(\nabla\times\mathbf{v})-(\mathbf{p}_2\cdot\nabla)\mathbf{v}.
\end{eqnarray}
Here, the last two terms of right side of Eq. (62) can be neglected
by Eq. (50a). Hence, we have
\begin{equation}
   \frac{d\mathbf{p}_2}{dt}=\frac{\partial\mathbf{p}_2}{\partial t}
   +\nabla(\mathbf{v}\cdot\mathbf{p}_2)-\mathbf{v}\times(\nabla\times\mathbf{p}_2).
\end{equation}

Inserting Eqs. (60) and (63) into Eq. (55), the acceleration of
selected point of test object can be written by
\begin{subequations}
\begin{equation}
   \mathbf{a}=-\frac{Q}{\phi_{10}\gamma^3_1}
   \left[\frac{\partial\mathbf{p}_2}{\partial t}
   -\mathbf{v}\times(\nabla\times\mathbf{p}_2)\right]+\mathbf{O},
\end{equation}
where
\begin{equation}
  \mathbf{O}=\left(\frac{1}{\phi_{10}\gamma^3_1}\right)
   \left[\mathbf{X}-Q\nabla(\mathbf{v}\cdot\mathbf{p}_2)\right]
   -\frac{1}{c^2}\mathbf{v}\times(\mathbf{v}\times\mathbf{a}).
\end{equation}
\end{subequations}

Next, substituting the constant of motion M for $\gamma_1$ of Eq.
(64a) using Eq. (51), we have
\begin{equation}
   \mathbf{a}=-\frac{Q}{M^3 \phi_{10}\psi^3_2}
   \left[\frac{\partial\mathbf{p}_2}{\partial t}
        -\mathbf{v}\times(\nabla\times\mathbf{p}_2)\right] +\mathbf{O}.
\end{equation}
Here, since the variable $\mathbf a$ in Eq. (65) is the acceleration
of selected point of test object caused by the movement of source
object, it corresponds to the afore-defined kinetic acceleration
$\mathbf{a}_k$. Therefore, inserting Eqs. (53) and (65) into Eq.
(52), the total acceleration of selected point of test object
$\mathbf{a}_t$, which is caused by the moving source object with a
non-uniform density distribution, can be written by

\begin{widetext}
\begin{eqnarray}
   \mathbf{a}_{t}=\frac{q_1 c^2}{M^2}\left(\frac{1}{\psi^3_2}\right)
   \left\{\nabla\psi_2
   - \frac{Q}{q_1 M \phi_{10} c^2}\left[\frac{\partial\mathbf{p_2}}{\partial t}
   -\mathbf{v}\times(\nabla\times\mathbf{p_2})\right]\right\}
   + \mathbf{O}.
\end{eqnarray}
Though similar, this equation is not identical with the known
Lorentz force formula, because of the non-constant factor
$1/\psi^3_2$. But, we can overcome this difference using the
following identities.
\begin{subequations}
\begin{eqnarray}
   \frac{1}{\psi^3}\nabla\psi
     &=&-\frac{1}{2}\nabla\frac{1}{\psi^{2}},\\
   \frac{1}{\psi^3}\nabla\times(\psi\mathbf{u})
     &=&-\frac{1}{2}\nabla\times\left(\frac{\mathbf{u}}{\psi^{2}}\right)
      +\frac{3}{2\psi^{2}}\nabla\times\mathbf{u},\\
   \frac{1}{\psi^3}\frac{\partial}{\partial t}(\psi\mathbf{u})
     &=&-\frac{1}{2}\frac{\partial}{\partial
     t}\left(\frac{\mathbf{u}}{\psi^{2}}\right)
      +\frac{3}{2\psi^{2}}\frac{\partial \mathbf{u}}{\partial t},\\
   \frac{1}{\psi^3}\frac{\partial{\psi}}{\partial t}
     &=&-\frac{1}{2}\frac{\partial}{\partial
     t}\left(\frac{1}{\psi^{2}}\right),\\
   \frac{1}{\psi^3}\nabla\cdot(\psi\mathbf{u})
     &=&-\frac{1}{2}\nabla\cdot\left(\frac{\mathbf{u}}{\psi^{2}}\right)
      +\frac{3}{2\psi^{2}}\nabla\cdot\mathbf{u}.
\end{eqnarray}
\end{subequations}
That is, using Eqs. (67a)-(67c), Eq. (66) can be re-written by
\begin{equation}
   \mathbf{a}_{t}= \frac{q_1 c^2}{2M^2}
   \left\{-\nabla\left(\frac{1}{\psi_2^2}\right)
   + \frac{Q}{q_1 M \phi_{10} c^2}\left[\frac{\partial}{\partial
     t}\left(\frac{\mathbf{u}}{\psi_2^{2}}\right)
   -\mathbf{v}\times\nabla\times\left(\frac{\mathbf{u}}{\psi_2^{2}}\right)\right]\right\}
   + \mathbf{O'},
\end{equation}
where
\begin{equation}
\mathbf{O}'=
\mathbf{O}-\frac{3Q}{2M^3\phi_{10}}\left(\frac{1}{\psi_2^2}\right)
                \left[\frac{\partial\mathbf{u}}{\partial t}
                -\mathbf{v}\times(\nabla\times\mathbf{u})\right].\tag{68a}
\end{equation}
\end{widetext}
At this time, the known Lorentz force formula can be obtained
completely from the terms in the brackets of Eq. (68), as will be
shown below. In this sense, if the classical electromagnetic theory
is exact, all of additional terms written in Eq. (68a) or the sum of
them should be vanished. Particularly, we can neglect the last term
of right side of Eq. (68a) in most classical cases, because of Eq.
(50b). But, if not, it is expected that we can measure physical
effects caused by the terms written in Eq. (68a), that is,
discrepancies from the known electromagnetic theory.

Now, let us derive the Lorentz force formula and the Maxwell
Equations. For this, let us assume that the term $\mathbf O'$ of Eq.
(68a) can be neglected and that $Q/q_1$, which depends on the signs
of electric charges of objects, is equal to -1, as a matter of
convenience. (Further study is needed to verify the properness of
these assumptions.) In addition, let us define a scalar function and
a vector function as follows:
\begin{subequations}
\begin{eqnarray}
\Phi       &\equiv& \frac{c^2}{2\psi^{2}_2},\\
\mathbf{A} &\equiv& \frac{\mathbf{u}}{2\psi^2_2}=
                   \Phi\frac{\mathbf{u}}{c^2}.
\end{eqnarray}
\end{subequations}
Then, the equation (68) can be simply expressed by
\begin{eqnarray}
   \mathbf{a}_{t}=\frac{q_1}{M^2}
   \left\{ -\nabla\Phi + \frac{1}{ M \phi_{10}} \left[\frac{\partial\mathbf{A}}{\partial t}- \mathbf{v}\times(\nabla\times\mathbf{A})\right]\right\}
\end{eqnarray}
Eq. (70) is identical with the Lorentz force except the additional
factors $M^2$ and $M^3\phi_{10}$, which denote respectively the
electric and magnetic susceptibilities as explained below. In this
sense, the scalar and vector functions of Eqs. (69a) and (69b) can
be understood as the scalar and vector potentials, respectively, in
the classical electromagnetic theory, and the electric field
\textbf{E} and the magnetic field \textbf{B} can be defined as
follows:
\begin{eqnarray}
   \mathbf{E}&=& \frac{1}{M^2}
     \left[-\nabla\Phi-\frac{1}{M \phi_{10}}\frac{\partial\mathbf{A}}{\partial t}\right],\\
   \mathbf{B}&=& \frac{1}{M^2}
     \left[\frac{1}{M \phi_{10}}\nabla\times\mathbf{A}\right].
\end{eqnarray}

Then, the divergences and curls of \textbf{E} and \textbf{B}-fields
are respectively given by
\begin{eqnarray}
   \nabla\cdot\mathbf{E}&=& \frac{1}{M^2}\left[-\nabla^2\Phi
     -\frac{1}{M \phi_{10}}\frac{\partial(\nabla\cdot\mathbf{A})}{\partial t}\right], \\
   \nabla\cdot\mathbf{B}&=&0,\\
   \nabla\times\mathbf{E}&=&-\frac{\partial \mathbf{B}}{\partial t},\\
   \nabla\times\mathbf{B}&=&\frac{1}{M^3 \phi_{10}}\left[\nabla(\nabla\cdot\mathbf{A})-\nabla^2\mathbf{A}\right].
\end{eqnarray}

Meanwhile, from the previous discussion, the physical world must be
understood as a continuum of ex-entity. In this sense, though not
mentioned up to now, the continuity equation must be the most
important and essential law for describing the physical world. The
continuity equation for the source object can be written by
\begin{equation}
   \frac{\partial \psi_2}{\partial t} +\nabla\cdot\mathbf{p}_2=0.
\end{equation}
Multiplying Eq. (77) by $c^2\psi^{-3}_2$ and using the above
identities (67d) and (67e), we have
\begin{equation}
   \left[ \frac{\partial}{\partial t}\left(\frac{c^2}{2\psi^2}\right)
   +\nabla\cdot\left(\frac{\mathbf{u}c^2}{2\psi^2}\right)\right]
   -\frac{3c^2}{2\psi^2}\nabla\cdot\mathbf{u}=0.
\end{equation}
Using the scalar potential $\Phi$ of Eq. (69a) and the vector
potential \textbf{A} of Eq. (69b), Eq. (78) can be re-written by
\begin{equation}
   \left[\frac{\partial \Phi}{\partial t} +c^2(\nabla\cdot\mathbf{A})\right]
   -3\Phi(\nabla\cdot\mathbf{u})=0.
\end{equation}
Owing to Eq. (53a), we can also neglect the last term of Eq. (79).
As a result, Eq. (79) can be re-written by
\begin{equation}
   \frac{\partial \Phi}{\partial t} +c^2(\nabla\cdot\mathbf{A})=0.
\end{equation}
This equation coincides with the "Lorentz condition". (It is
interesting that the Lorentz condition can be obtained from the
continuity equation.) Using Eqs. (71) and (80), the above Eqs. (73)
and (76) can be written respectively by
\begin{eqnarray}
   \nabla\cdot\left(M^2\mathbf{E}\right)&=&
   \frac{1}{M\phi_{10}c^2} \frac{\partial^2\Phi}{\partial t^2}-\nabla^2\Phi,\\
  \nabla\times\left(M^3\phi_{10}\mathbf{B}\right)&=&
  \frac{1}{c^2}\frac{\partial \left(M^2\mathbf{E}\right)}{\partial
  t}\nonumber\\
  &+&\left[\frac{1}{M\phi_{10}c^2}\frac{\partial^2\mathbf{A}}{\partial t^2}-\nabla^2\mathbf{A}\right].
\end{eqnarray}

If the scalar and the vector potentials satisfy the following wave
equations,
\begin{eqnarray}
   \frac{1}{M\phi_{10}c^2}\frac{\partial^2\Phi}{\partial t^2}-\nabla^2\Phi
   &=&\frac{\rho}{\epsilon_0},\\
   \frac{1}{M\phi_{10}c^2}\frac{\partial^2\mathbf{A}}{\partial t^2}- \nabla^2\mathbf{A}
   &=&\frac{\mathbf{j}}{\epsilon_0 c^2}.
\end{eqnarray}
the above equations (81) and (82) can be written in the simple form
as follows:
\begin{eqnarray}
  \nabla\cdot\left(M^2\epsilon_0\mathbf{E}\right)&=&\rho,\\
  \nabla\times\left(M^3\phi_{10}\epsilon_0 c^2\mathbf{B}\right)&=&
  \frac{\partial }{\partial t}\left(M^2\epsilon_0\mathbf{E}\right)+\mathbf{j}~.
\end{eqnarray}
(The properness of above wave equations is also needed to study
furthermore.) In addition, if we introduce the famous Maxwell
relation $c^2=1/{\epsilon_0\mu_0}$ and define the electric
displacement \textbf{D} and the magnetic intensity \textbf{H} as
follows:
\begin{eqnarray}
  \mathbf{D}&\equiv& M^2\epsilon_0\mathbf{E}=(1+\chi_e)\epsilon_0\mathbf{E},\\
  \mathbf{H}&\equiv& \frac{M^3\phi_{10}}{\mu_0}\mathbf{B}=\frac{\mathbf{B}}{({1+\chi_m})\mu_0},
\end{eqnarray}
the above equations (85) and (86) can be written by
\begin{eqnarray}
  \nabla\cdot\mathbf{D}&=&\rho,\\
  \nabla\times\mathbf{H}&=&\frac{\partial \mathbf{D}}{\partial t}+\mathbf{j}.
\end{eqnarray}
Equations (74), (75), (89) and (90) are identical with the known
Maxwell equations: Maxwell equations are derived.

In addition, comparing the above Eqs. (87) and (88) with the known
formulae between \textbf{E}, \textbf{D}, \textbf{B} and \textbf{H},
we have
\begin{eqnarray}
M^2&=& 1+\chi_e = k_e,\\
M^3\phi_{10}&=& \frac{1}{1+\chi_m}= k_m.
\end{eqnarray}
where $\chi_e$ and $\chi_m$ denote the electric and magnetic
susceptibilities respectively, and $k_e$ and $k_m$ denote the
dielectric constant and the relative permeability respectively.
Given that the constant of motion M means (total mechanical
energy)/$m_0c^2$ as discussed in the sections III.E.2-3, we can
correlate $\chi_e$ and $k_e$ with the total mechanical energy by
using Eq. (91). Though not discussed here, this issue is also needed
to study furthermore.

Furthermore, from Eqs. (91) and (92), we can have
\begin{equation}
  M\phi_{10} c^2=\frac{1}{\epsilon\mu} \equiv c^2_m,
\end{equation}
where
\begin{equation}
  c^2_m = \frac{c^2}{(1+\chi_e)(1+\chi_m)},
\end{equation}
where $\epsilon$ and $\mu$ denote the permittivity and the
permeability at the position of selected point. Using Eq. (93), the
above wave equations (83) and (84) are given by
\begin{eqnarray}
   \frac{1}{c^2_m}\frac{\partial^2\Phi}{\partial t^2}-\nabla^2\Phi
   &=&\frac{\rho}{\epsilon_0},\\
   \frac{1}{c^2_m}\frac{\partial^2\mathbf{A}}{\partial t^2}- \nabla^2\mathbf{A}
   &=&\frac{\mathbf{j}}{\epsilon_0 c^2}.
\end{eqnarray}

Finally, from the above Eqs. (54b) and (69a), the scalar potential
$\Phi$ can be expressed by using Eqs. (19) and (42), as follows:
\begin{eqnarray}
   \Phi=\frac{c^2}{2}\left(1-\frac{u^2}{c^2}\right)\left(1+\frac{2q_s k_s}{rc^2}\right).
\end{eqnarray}
Here, for $u/c\ll 1$, Eq. (97) can be approximately written by
\begin{eqnarray}
   \Phi\approx\frac{c^2}{2}+\frac{q_s k_s}{r}.
\end{eqnarray}
Since the term $c^2/2$ of Eq. (98) is constant, it plays no role in
the electromagnetic laws written as the differential form. In this
sense, we may say that the classical Coulomb potential is
approximation of Eq. (97) for the case of slow source object,
because $k_s$ is $1/4\pi \epsilon_0$ as mentioned above.
Nevertheless, the above equation (97) has some difference with the
known relativistic scalar potential\cite{Feynman4}. As a result, the
vector potential \textbf{A} has also some difference with the known
relativistic vector potential, because the vector potential
\textbf{A} is defined by the scalar potential $\Phi$ as seen in the
Eq. (69b). I hope that further studies will be pursued for
overcoming this difference and for completing the afore-mentioned
some assumptions.

\subsubsection{Quantum Mechanics}
From the previous discussions, matter and space is classified based
on the difference in the density of ex-entity, and the movement of
universally distributed object can be understood as the phenomenon
of wave-like propagation thereof. In this sense, it seems that wave
and particle are not contradictory concepts but only concepts that
fall under different categories. That is, the wave corresponds to
the phenomenal concept that represents the change mode of ex-entity
or the physical phenomena, and the particle does the concept on
state that represents the quantitative state of ex-entity or the
distributional state of ex-entity density that has strong stability.
Hence, the duality of matter and wave, which was the starting point
of quantum mechanics, is not contradictory.

In addition, the function of probability density in the
Shr\"{o}dinger equation is very similar to the afore-discussed
function of ex-entity density in that they contain all physical
information of system. In this sense, it is obvious that the
function of probability density should be understood in connection
with the function of ex-entity density, even though we cannot say
that two density functions are identical with each other. This
subject is also needed to study furthermore. Additionally, in this
section III.F., we have mentioned several subjects required for
further study, but it seems that most of these subjects are closely
related to the quantum mechanics.

\section{Acknowledgement}
I would like to thank the late Mr. Albert Einstein, because his
achievement in the relativity theory was a starting point of this
work and has been a compass that has guided me to the above results.
And, I am indebted to my wife, Choi, Yu Jung, for economical support
and for helpful advice during the translation of manuscript.

\end{document}